\documentclass[trackchanges,twocolumn,nofloatfix]{aastex7}

\usepackage{xspace}
\usepackage{siunitx}
\usepackage{shortbold}
\usepackage{hyperref}
\usepackage{amsmath}
\usepackage{comment}
\usepackage{booktabs}
\usepackage{pgfplotstable}
\usepackage[table]{xcolor}
\usepackage{soul}
\usepackage{placeins}

\definecolor{lightgray}{gray}{0.9}

\DeclareSIUnit{\DN}{DN}
\newcommand{\DNS}{\SI{}{\DN\per\second}}
\newcommand{\coloneqq}{\mathrel{\mathop:}=}

\newcommand{\abr}{$\alpha B_R$\xspace}
\newcommand{\abp}{$\alpha B_\phi$\xspace}
\newcommand{\abt}{$\alpha B_\theta$\xspace}

\newcommand{\gauss}{Mx\,cm$^{-2}$\xspace}
\newcommand{\gpx}{\gauss\,\ensuremath{\mathrm{pixel}^{-1}}}

\newcommand{\logvar}{\log\boldsymbol{\sigma}^{2}}

\newcommand{\modelname}{\text{MAGiDiff}\xspace}

\newcommand{\aia}{{{\it SDO}/AIA}\xspace}
\newcommand{\stereo}{{\it STEREO}/EUVI\xspace}
\newcommand{\stereoA}{{\it STEREO-A}/EUVI\xspace}
\newcommand{\goes}{{\it GOES}/SUVI\xspace}
\newcommand{\goessixteen}{{\it GOES--16}/SUVI\xspace}
\newcommand{\goeseighteen}{{\it GOES--18}/SUVI\xspace}

\newcommand{\hinode}{{\it Hinode}/SOT-SP\xspace}

\newcommand{\hmi}{{{\it SDO}/HMI}\xspace}

\newcommand{\suvi}{{{\it GOES-R}/SUVI}\xspace}

\newcommand{\AAs}{\AA\xspace}
\newcommand{\dnpersec}{DN s$^{-1}$\xspace}

\begin{document}

\title{\modelname: Sampling the Photospheric Vector Field from UV/EUV Filtergrams}

\author[0009-0007-7942-826X]{Ruoyu Wang}
\affiliation{New York University, Courant Institute of Mathematical Sciences}
\email{rw3544@nyu.edu}

\author[0000-0001-5028-5161]{David Fouhey}
\affiliation{New York University, Courant Institute of Mathematical Sciences}
\affiliation{New York University, Tandon School of Engineering}
\email{dff6142@nyu.edu}

\begin{abstract}
Photospheric vector magnetic fields are foundational to modeling, understanding, and forecasting solar activity. These data are usually produced by inverting and disambiguating the full Stokes vector at multiple passbands, which is demanding. Here, we investigate how well we can estimate photospheric vector magnetograms from UV/EUV filtergrams. This problem is challenging and intrinsically ambiguous without polarization information, as the mapping from UV/EUV intensity to the magnetic field is indirect and ill-posed. We introduce \modelname, a machine-learning-based method that uses denoising diffusion models to estimate vector magnetograms from UV/EUV filtergrams. As input, \modelname takes a stack of filtergrams from the Solar Dynamics Observatory (SDO) / Atmospheric Imaging Assembly (AIA); as output, it is trained to estimate the disambiguated vector magnetogram as seen by Hinode / Solar Optical Telescope-Spectro-Polarimeter (SOT-SP). We show that \modelname can accurately mimic the Hinode ground truth. Additionally, we probe \modelname's  understanding of the physical structure and magnetic connectivity. On full-disk, we show that it produces plausible structures for active regions. \modelname generalizes across solar cycles despite hemispheric polarity reversal, and can be fine-tuned to other EUV instruments including \stereo and \suvi. While clearly not a substitute for a dedicated instrument, \modelname opens the door to new capabilities.

\end{abstract}

\section{Introduction}
\label{sec:intro}
In solar physics, high-quality photospheric vector magnetic field measurements underpin our understanding of many solar activities, including space weather forecasting \citep{bobra2014helioseismic, Bobraetal2015, barnes2016comparison, DAFFS}, coronal magnetic structure modeling and magnetohydrodynamics (MHD) simulation \citep{wiegelmann2021solar, JIANG2022100236}, and the evolution of the Sun's atmosphere \citep{LekaBarnes2007, cheung2012method, Fisher2012, Lionello2014, gombosi2018extended, hayashi2021coupling, schuck2022origin}. 
However, estimating vector magnetograms usually requires sampling the full Stokes vector (i.e., [$I,Q,U,V$]) at multiple wavelengths. These signals are difficult to acquire, especially the linear polarization components $Q$ and $U$~\citep{Stenflo94}. 
As a result, vector magnetograms can be limited in cadence, coverage, or availability across missions. This motivates a complementary approach: evaluating how well vector magnetograms can be estimated from other more widely available observations. Such estimates cannot replace spectropolarimetric measurements, but a reliable estimate would extend the use of existing data archives and provide magnetic field context in settings where direct vector measurements are not available. 

UV/EUV intensity maps are a natural starting point because they are abundant and are routinely collected by a broader set of instruments, many of which do not carry spectropolarimeters. Instruments like the Atmospheric Imaging Assembly (AIA; \cite{lemen_atmospheric_2012}) on SDO, the Extreme Ultraviolet Imager (EUVI; \cite{Wuelser_EUVI_2004}) on the twin Solar Terrestrial Relationship Observatory (STEREO; \cite{kaiser_stereo_2008}), the Extreme Ultraviolet Imager (EUI; \cite{Rochus_EUI_SO_2020}) on Solar Orbiter (SO; \cite{Muller_SO_2020}), and the Solar UltraViolet Imager (SUVI; \cite{darnel_goes-r_2022}) on the Geostationary Operational Environmental Satellites (GOES-R series; \cite{Hill_GOES_satellite_2005}) deliver high-cadence, full-disk, multi-wavelength observations of the corona over decades. 

Unfortunately, there is a fundamentally ambiguous and statistical relationship between UV/EUV intensity and the photospheric magnetic field. First, if one were to flip the field consistently, one would observe identical intensity images (in addition to the $180^\circ$ plane-of-sky ambiguity described by~\cite{Harvey69}). Second, most of the UV/EUV light does not originate at the photosphere. 
Yet, while the relationship is less direct than in Stokes inversion, there is signal. For instance, the known statistical relationship between unsigned flux and intensity~\citep{schrijver1987solar} has been used to derive total flux from 304\AAs data \citep{ugarte2015magnetic,knizhnik2024effects}. Similarly, due to the low plasma-$\beta$ of the corona, filtergrams originating in the corona (e.g., 171\AA) trace out magnetic field lines that head down to the photosphere. However, each observation provides an indirect cue that needs to be integrated to constrain the possible photospheric field.

While the polarity ambiguity is intrinsic, it can be empirically mitigated with additional context. Hale's law~\citep{Hale1925} states that the leading polarity of bipolar active regions is opposite in the northern and southern hemispheres, and this pattern reverses between consecutive solar cycles. Given heliographic latitude and solar cycle information as input, a model can often infer the polarity without the need for polarization measurements.  

These observations motivate \modelname, a learning-based framework that maps UV/EUV filtergrams to heliographic vector magnetic flux density components \abr, \abp, and \abt (where the $\alpha$ denotes that the quantity includes both intrinsic field strength and filling fraction $\alpha$). \modelname takes input from \aia, due to its  mission archive of 15+ years of nearly-complete, high-quality, filtergrams. \modelname is built on denoising diffusion models, a recent advance in machine learning that has produced remarkable results in conditional image generation. Unlike regression models that predict a single estimate, diffusion models learn to sample from a distribution of possible predictions, producing sharper outputs and enabling uncertainty estimation.
\modelname is trained to estimate magnetic flux density from \hinode observations resampled to the \hmi sampling grid. One reason for using \hinode is entirely practical: its dense sampling of two absorption lines yields high-quality magnetograms, which enhances the overall quality of \modelname's output. The other is forward-looking: \hinode captures only a limited field of view, but many sources of high-quality data will also be non-full disk, like PHI~\citep{Solanki_PHI_SO_2020} for magnetograms outside the Sun-Earth line.  

We evaluate \modelname on several tests. First, we show its performance on previously unseen regions, including across solar cycles. Second, we probe \modelname's ability to model the correlations in the vector field by analyzing its predicted distribution. Third, we test the model's generalization ability on full-disk inputs. Finally, we show that \modelname can be fine-tuned to other instruments like \stereo, \goessixteen, and \goeseighteen, demonstrating a path toward supplemental vector field estimation from missions without spectropolarimeters. While not a substitute for a dedicated facility that captures spectropolarimetric data and inverts it to produce vector magnetograms, we believe that \modelname presents an opportunity to get more from existing data.

\section{Data} \label{sec:data}
This work tackles the task of estimating possible \hinode vector magnetograms from UV/EUV data obtained from \aia. A full description of the data involved is beyond the scope of the paper, but we give a brief description to make the paper more self-contained.

The input data from \aia consist of high-cadence, high-resolution filtergram observations, grouped in EUV (94\AA, 131\AA, 171\AA, 193\AA, 211\AA, 304\AA, 335\AA) and UV (1600\AA, 1700\AA) filters. We do not use \aia's visible light observation (at 4500\AA). Each filter has a characteristic temperature and height at which much of the light is emitted.
These filtergrams are captured near-synchronously at $4096 \times 4096$ with $0.6\arcsec$ sampling and approximately $1.5\arcsec$ optical resolution. We use the {\tt aia.lev1\_euv\_12s} and {\tt aia.lev1\_uv\_24s} series, corrected for exposure time and filter degradation following~\cite{galvez2019machine}.

The output data are from \hinode, a scanning-slit spectropolarimeter that sweeps a slit across a scene, capturing dense Stokes profiles of two photospheric spectral lines at a spatial sampling of $0.16 \arcsec$. Each scan covers roughly $160 \arcsec$ along the slit with varying scan width, and typically takes about an hour to complete. This dense sampling yields higher-quality vector magnetograms than those from \hmi. The vector magnetic field is estimated from these data with a Stokes inversion technique~\citep{Lites_MERLIN_2007}, followed by ME0 disambiguation~\citep{Metcalf_ME0_1994} to yield the $180^\circ$-disambiguated~\citep{Harvey69} heliographic components.

The data are prepared as a co-aligned cutout volume of \aia filtergrams and \hinode vector magnetograms. We co-align to the frame of the Helioseismic and Magnetic Imager, HMI~\citep{schou_design_2012} to take advantage of the co-alignments of \hinode and \hmi done by~\cite{wang_supersynthia_2024} that account for a pointing error in \hinode data~\citep{Fouhey_Large-scale_2023}. We likewise align the \aia data to this grid. While \hinode takes tens of minutes to acquire a scan (resulting in intrinsic temporal misalignment), we let the method handle potential mismatch rather than handle it in data alignment.
In addition to the UV/EUV filtergrams, we provide a \textit{polarity prior} consisting of per-pixel heliographic latitude and a solar cycle indicator ($1$ for odd cycles and $-1$ for even cycles) as auxiliary inputs, both derived from the \aia FITS headers. This prior supplies the positional and temporal context needed to leverage Hale's law for polarity inference. 

The \hinode observations in our dataset are unfiltered, including active regions, plage, and quiet Sun.
We obtain 71,243 pairs of co-registered \aia intensity images and \hinode vector magnetograms spanning from 2011 to 2024. Following \cite{wang_supersynthia_2024}, data are split by acquisition time: the test set consists of years 2016 and 2024, with half-year buffers on both sides (2015 July - December, 2017 January - June and 2023 July - December) omitted to prevent data leakage from slowly evolving solar structures. The validation set consists of $2015$ January - June, $2017$ July - December and $2023$ January - June, and the remaining data form the training set. After filtering pairs with too small magnetograms to yield a $128\arcsec \times 128\arcsec$ crop, the training, validation, and test sets consists of $29112$, $6137$, and $8307$ samples, respectively.

We also prepare fine-tuning datasets for generalization to \stereo and \suvi. These EUV filtergrams are paired with SuperSynthIA~\citep{wang_supersynthia_2024} vector magnetograms (a ML-based method that produces \hinode-like magnetograms from \hmi data) reprojected to the corresponding instrument's coordinate grid. A full description of the reprojection, calibration and normalization details appears in Appendix \ref{sec:appx_cross_instr_fine_tuning}.

\section{Method} \label{sec:method}
We formulate the task of predicting solar vector magnetograms 
from UV/EUV intensity observations as {\it sampling} a photospheric vector magnetogram conditioned on a stack of contemporaneous UV/EUV maps. Let $\IB \in \mathbb{R}^{H \times W \times C} $ denote~ the conditioning information, consisting primarily of a stack of co-registered UV/EUV filtergrams, together with auxiliary inputs that provide physical context.
Let $\BB \in \mathbb{R}^{H \times W \times 3}$ be the corresponding magnetogram components ($\alpha B_R$, $\alpha B_\phi$, $\alpha B_\theta$). Our goal is to learn a model that can sample $\BB_n$ from $p(\BB \mid \IB_{n})$ where $\IB_{n}$ is unseen in training. This formulation allows us to sample multiple plausible magnetograms given the same UV/EUV input.

\begin{figure*}[tp]
    \centering
    \includegraphics[width=1\linewidth]{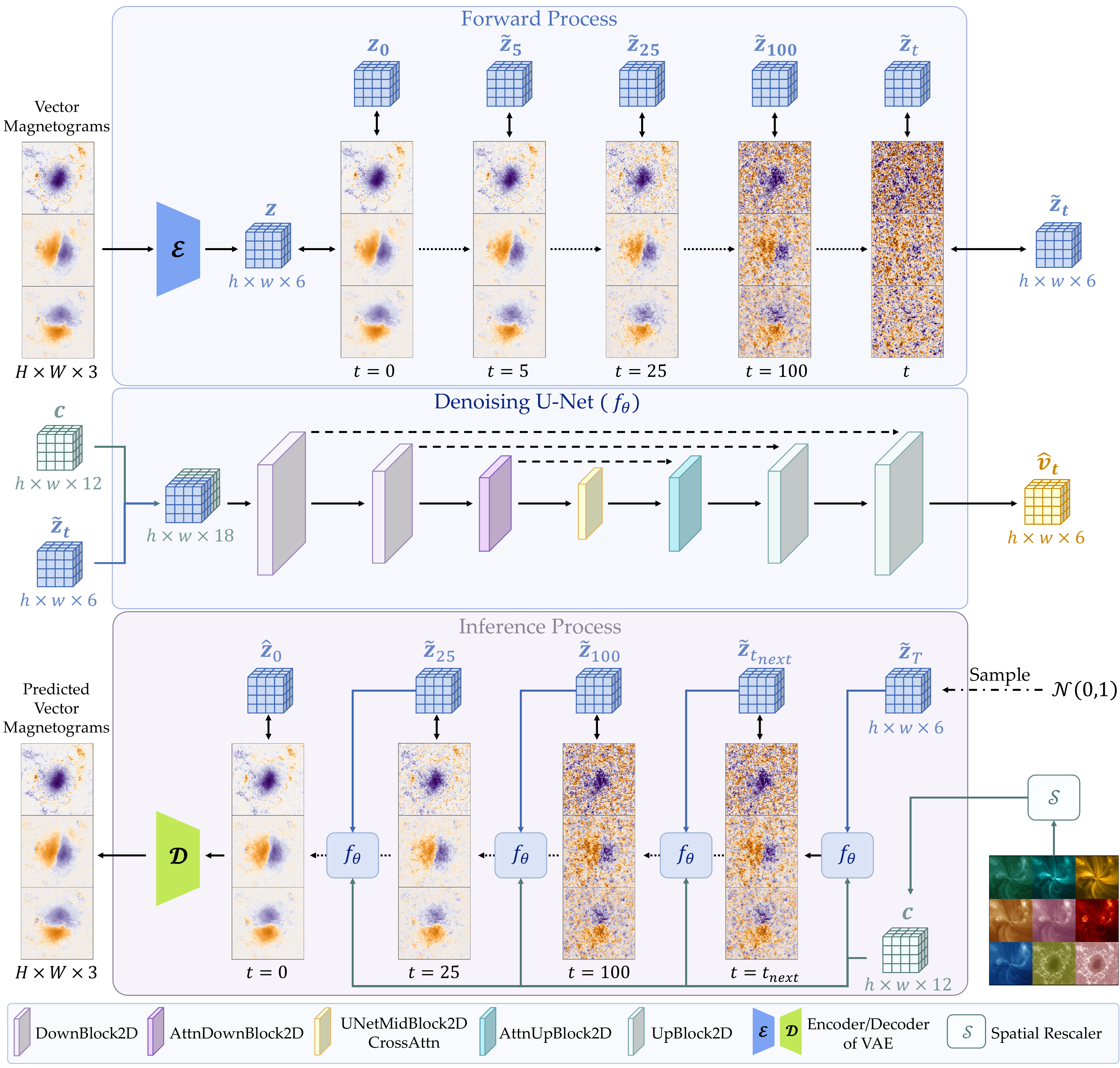}
    \caption{\textbf{ Architecture of denoising U-Net used in \modelname and illustration of the diffusion process.} 
      The denoising UNet takes noisy latent $\tilde{\zB}_t$ at timestep $t$ together with conditioning input $\cB$ as input and predicts the velocity $\hat{\vB}_t$ — a linear combination of $\epsilonB$ and $\xB_0$. In the forward process, the frozen VAE encoder maps an input magnetogram to a clean latent $\zB$, a timestep $t \in \{1, 2, \cdots, T \}$ is sampled, and the scheduler corrupts $\zB$ to $\tilde{\zB}_t$ according to its noise schedule. The UNet is trained to regress the corresponding velocity target. 
      At inference, sampling begins from $\tilde{\zB}_T$ sampled from Gaussian noise, and iterates the reverse process: at each step the UNet takes noisy latent at current timestep $\tilde{\zB}_t$ along with conditioning input $\cB$ and predicts velocity $\hat{\vB}_t$. The scheduler then uses $\hat{\vB}_t$ to remove noise from $\tilde{\zB}_t$. The final clean latent $\hat{\zB}_0$ is decoded by the frozen VAE decoder to produce the predicted vector magnetogram. Both the forward and reverse diffusion processes operate entirely in latent space. The intermediate states shown in the figure therefore correspond to latent variables, which are decoded by the frozen VAE decoder and displayed in magnetogram space only for visual interpretability.
     }
    \label{fig:architecture_LDM}
\end{figure*}

\begin{figure*}[t]
    \centering
    \includegraphics[width=1\linewidth]{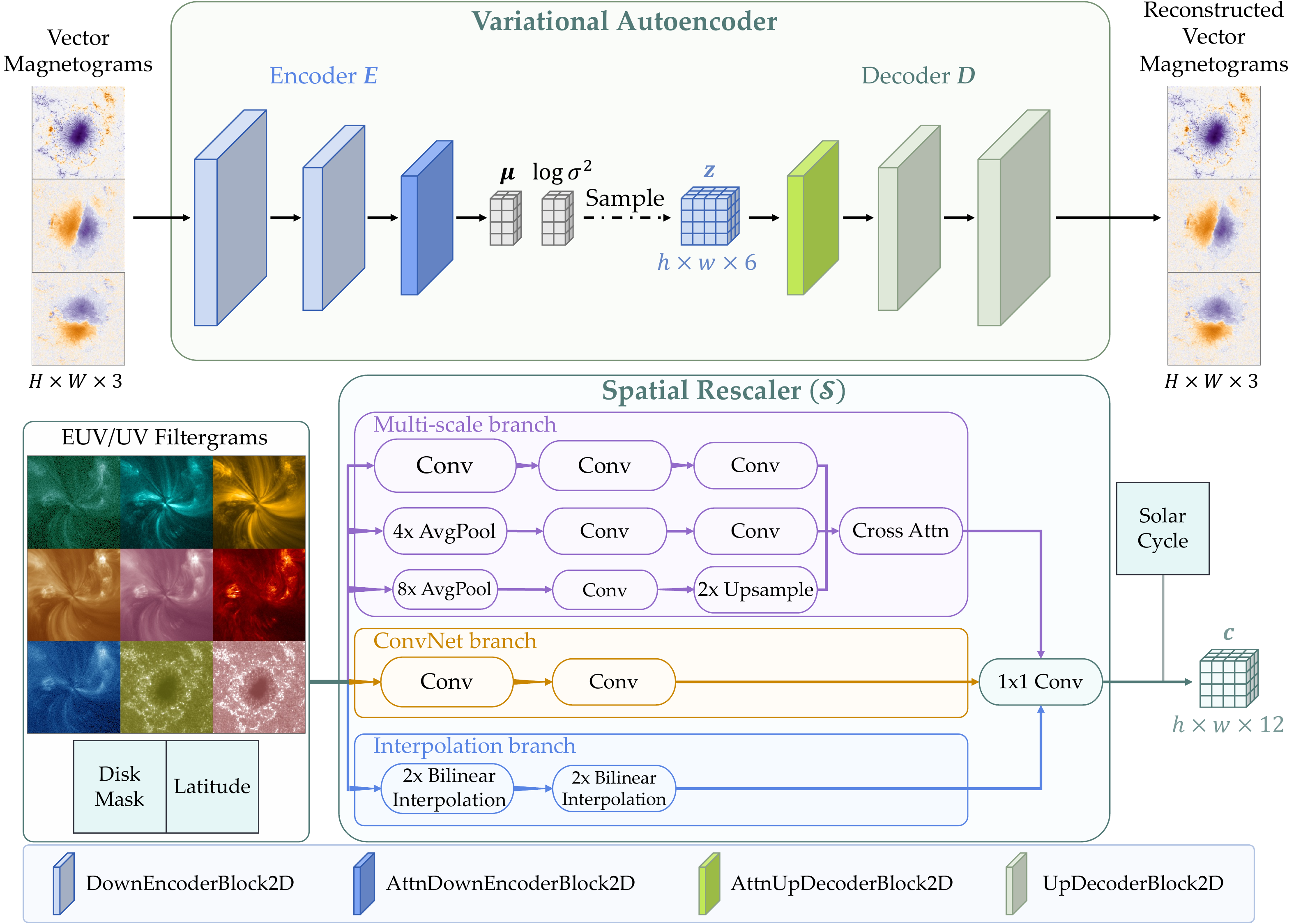}
    \caption{\textbf{ Architecture of Variational autoencoder and spatial rescaler used in \modelname.} 
      The VAE consists of an encoder and a decoder: the encoder maps three-component vector magnetograms to a gaussian latent distribution parameterized by mean $\mu$ and log-variance $\logvar$, from which a latent code $z$ is sampled; the decoder reconstructs the magnetogram from $z$. The spatial rescaler maps UV/EUV filtergrams—concatenated with auxiliary disk-mask and latitude channels—to a conditioning input $c$ at the latent's spatial resolution, via three parallel branches: a multi-scale convolutional branch with cross-scale attention fusion, a learnable convolution branch, and a bilinear interpolation branch. These branch outputs are concatenated channel-wise and merged by a $1 \times 1$ convolution. 
     }
    \label{fig:architecture_VAE}
\end{figure*}

\subsection{Diffusion Framework}
We use denoising diffusion probabilistic models (DDPMs; \cite{ho_denoising_2020}) to learn to model $p(\BB \mid \IB)$. 
These have shown efficacy in generating high fidelity images conditioned on other information, such as text. 
A full introduction to DDPMs is beyond the scope of the paper, but the reader is directed to the  introduction by \cite{yang2024_diffusion_survey}. 
Briefly, DDPMs consist of two phases: a \textit{forward} process that corrupts the image by adding Gaussian noise gradually over timesteps $1,\ldots,T$, and a \textit{reverse} process where a parametric model ($f_\theta$, where $\theta$ represents the model's parameters) is tasked to remove noise. 
By learning to remove this noise, the parametric model learns the given distribution.
\autoref{fig:architecture_LDM} provides an overview of the diffusion process used by \modelname.
For clarity, we use magnetogram-space notation in this high-level description of the DDPM framework. In the actual implementation of \modelname, the diffusion process operates entirely in latent space, as introduced below and detailed in Appendix~\ref{sec:appx_ldm_walkthrough}.
During training, given a data pair $(\IB, \BB)$, a timestep $t$ is randomly sampled $t \sim \mathrm{Uniform}\{1, \dots, T\}$ and a noisy magnetogram $\tilde{\BB}_t$ is calculated via \autoref{eq:forward_closed}. We fit the model $f_\theta$ to minimize an $\ell_2$ objective between the injected noise $\epsilonB$ and predicted noise $f_\theta (\tilde{\BB}_t, \IB, t)$: 
\begin{equation}
    \mathcal{L} = \mathbb{E }_{\BB_0, \epsilonB, t} \| \epsilonB - f_\theta (\tilde{\BB}_t, \IB, t) \|^{2}.
\end{equation}
During inference, we start with random gaussian noise $\tilde{\BB}_T \sim \mathcal{N}(0,1)$ and iteratively (for $t = T, T-1, \dots 1$) apply the learned denoiser $f_\theta$.

In practice, we use several recent innovations to DDPM to improve training stability and efficiency. A full description of technical details appears in Appendix \ref{sec:appx_diffusion_model_details}. Here, we briefly summarize their key steps.

\textbf{Latent Diffusion Model.} We adopt the latent diffusion model (LDM) structure \citep{rombach_high-resolution_2022}, where the forward and reverse processes are performed in a lower dimensional latent space, reducing spatial size while preserving image fidelity. The latent space is created by a variational autoencoder (VAE; \cite{VAE_Kingma_2013}) whose encoder and decoder translate between pixel and latent space representations.

\textbf{V-objective.} We use the v-objective \citep{v_obj_Salimans}, where the model predicts a velocity term $\vB$ (a linear combination of noise $\epsilonB$ and clean magnetogram $\BB_0$) rather than predicting noise $\epsilonB$ or clean magnetogram $\BB_0$ alone. This formulation stabilizes training and yields higher sample quality.

\textbf{Zero Terminal SNR.} We enforce zero terminal signal-to-noise ratio following \cite{zero_diffusion_snr_Lin_2024}, meaning that at timestep $T$ all structured signal are removed, leaving pure noise $\BB_T = \epsilonB$. This allows \modelname to estimate magnetograms with arbitrary mean intensity, capturing the full dynamic range of solar magnetic field strengths.

\textbf{DDIM Sampling.} During inference, we employ the denoising diffusion implicit model (DDIM; \citep{song_denoising_2022}) sampler, which replaces the original stochastic, Markov inference process with a non-Markov, deterministic update path. This approach enables generating high-fidelity magnetogram predictions in fewer steps.

\subsection{Model Architecture}
\modelname consists of three components: a VAE that maps between pixel space and latent space, a spatial rescaler that projects conditioning UV/EUV maps to latent dimension, and a denoising U-Net that estimates latent representation of the clean vector magnetogram. These components are built with the HuggingFace Diffuser library. 
\autoref{fig:architecture_LDM} illustrates the denoising U-Net structure, while  \autoref{fig:architecture_VAE} details the VAE and spatial rescaler architectures.

\textbf{VAE.} We adopt the AutoencoderKL from the HuggingFace Diffuser stable diffusion implementation. The encoder consists of two convolutional downsampling blocks followed by a self-attention downsampling block (DownEncoderBlock2D, DownEncoderBlock2D, AttnDownEncoderBlock2D), which progressively compress the input magnetograms into a compact latent representation. The decoder mirrors the structure to reconstruct latent back into pixel space. The latent representation is downsampled by a factor of $4$ in height and width and is set to have six channels, balancing reconstruction quality with computational efficiency.

\textbf{Spatial Rescaler for UV/EUV Filtergrams.} The conditioning UV/EUV filtergrams are encoded by a multi-branch neural network designed for efficient spatial downsampling while preserving contextual information across scales. It consists of three parallel processing branches: (1) a multi-scale learnable feature extraction branch that captures feature at 1/4 and 1/8 resolution and then fused through cross-attention, (2) a learnable ConvNet branch, and (3) a traditional interpolation branch. All branches are ultimately concatenated and projected to the desired output dimension. This design enables the module to maintain fine details while capturing broader contextual information during downsampling. It also allows the spatial rescaler to learn on its own which features are important to extract, while still benefiting from context provided by traditional interpolation method.

\textbf{Denoising U-Net.} The denoising backbone adopts UNet2DConditionModel, also from the stable diffusion implementation. The encoder of the denoising U-Net consists of two convolutional downsampling blocks followed by an self-attention downsampling block (DownBlock2D, DownBlock2D, AttnDownBlock2D), enabling the model to gradually reduce spatial dimensions while capturing global context through an attention mechanism. At the bottleneck, we use a UNetMidBlock2DCrossAttn module with no external conditioning. In this configuration, the cross-attention layers reduce to self-attention, allowing the model to capture long-range dependencies within the latent representation while keeping the architecture compatible with potential conditional information for future work. The decoder symmetrically upsamples and fuses features via skip connections. This architecture maintains a balance between local detail preservation and large-scale contextual modeling, which is essential for reconstructing complex magnetic field patterns.

\subsection{Data Normalization}
Machine learning systems generally prefer data to be well normalized for stable and efficient training. 
Gradient-based optimizers can struggle when feature values vary by orders of magnitude, which can lead to vanishing and exploding gradients. 
This is exactly the case for both the UV/EUV intensity maps and the vector magnetograms.

For \aia UV/EUV intensity data, quiet-Sun pixels are typically tens to hundreds counts per second (\SI{}{\DN\per\second}), whereas active region pixels can reach several thousand \SI{}{\DN\per\second}. 
To make this long-tail distribution more suitable for training, we apply a log1p transform, defined as $\log(1+x)$, and then standardize each \aia channel using its per-channel mean and standard deviation.

Apart from the UV/EUV filtergrams, the conditioning input includes auxiliary channels that are encoded separately. The disk mask is kept binary. The heliographic latitude map is expressed in degrees, divided by $90^\circ$ to map on-disk values to $[-1, 1]$; off-disk pixels are assigned a value of $-2$. The solar cycle indicator is encoded as $1$ for odd-numbered cycles and $-1$ for even-numbered cycles. We therefore do not apply additional normalization to these auxiliary channels.

Although less pronounced than the \aia intensity maps, vector magnetograms are also by nature long-tailed: most pixel values are close to zero, while active-region fields can reach several thousand \gauss. 
For each vector magnetic field component, we divide the absolute field strength by $4000$~\gauss, take the square root, and restore the original sign. Thus, a $4000$~\gauss~ component has unit magnitude after normalization.

\subsection{Training and Inference Process}
\label{sec:method-Training_and_Inference_Process}
We summarize the notation and workflow of \modelname for clarity. The full model is denoted $F$ with the denoising U-Net $f_\theta$ parameterized by learnable weights $\theta$. 
The Variational Autoencoder (VAE) is represented with the encoder-decoder pair ($\mathcal{E}$, $\mathcal{D}$), and the spatial rescaler for encoding conditional UV/EUV filtergrams is denoted as $\mathcal{S}$.
The conditional information, denoted $\IB \in \mathbb{R}^{H \times W \times 12}$ as described before, is a concatenation of four types of input: (1) nine co-registered \aia UV/EUV filtergrams; (2) a binary disk mask indicating whether each pixel lies on the solar disk; (3) per-pixel heliographic latitude, computed from the \aia WCS information; (4) a solar cycle indicator, with value $1$ for odd-numbered cycles and $-1$ for even-numbered cycles.
The target vector magnetogram is written as $\BB \in \mathbb{R}^{H \times W \times 3}$ with the three heliographic vector magnetic field components:  $\alpha B_\phi$ (longitudinal); $\alpha B_\theta$ (latitudinal); $\alpha B_R$ (radial). 
We use ($h$, $w$) to denote the spatial dimension of latents for input with shape ($H$, $W$). In our setup, $h = H/4$, $w = W/4$, and the latent dimension is set to $6$. We denote~ $\hat{}$ ~as prediction,~$\tilde{}$ ~as noisy.

\subsubsection{Training of VAE}

\modelname training begins with the training of a VAE from scratch on $128\arcsec \times 128\arcsec$ \hinode vector magnetogram cutouts. Given input magnetogram $\BB$, encoder $\mathcal{E}$ maps it to parameters for a Gaussian distribution in latent space, producing mean $\muB$ and log-variance $\logvar$:
\begin{equation}
    (\muB, \logvar) \coloneqq \mathcal{E}(\BB)
\end{equation}
where the actual variance $\boldsymbol{\sigma}^{2}$ is essentially $\boldsymbol{\sigma}^{2} = \exp(\logvar)$. The Gaussian posterior can thus be expressed as:
\begin{equation}
    q(\zB \mid \BB) = \mathcal{N}(\zB; \muB, \sigmaB^{2} \IB)
\end{equation}
from which we can sample latent via reparameterization as $\zB = \muB + \sigmaB \odot \epsilonB$ where $\epsilonB \sim \mathcal{N}(0,1)$. The decoder $\mathcal{D}$ then maps the latent back and produces reconstructed magnetogram $\hat{\BB} = \mathcal{D}(\zB)$. $\ell_1$ and Kullback-Leibler divergence (KL-Divergence; \cite{kl_div_Kullback_1951}) loss are used to supervise the training process. The total loss is thus: 
\begin{equation}
    \mathcal{L}(\BB) = \| \BB - \hat{\BB} \|_{1} + \lambda D_{KL}(q(\zB \mid \BB)~\|~ \mathcal{N}(0,\IB))
\end{equation}
where $\lambda$ controls the extent to which we regularize the latent distribution and is set to $10^{-3}$, encouraging faithful reconstruction while maintaining well regularized latent space. We found that insufficient KL regularization can leave the VAE reconstructions visually accurate while producing a latent space that is more difficult for the subsequent diffusion model training, leading to spurious artifacts in the predicted magnetograms particularly in quiet-Sun. Conversely, excessive regularization overly constrains the latent space, causing the predicted magnetograms to lose fine-scale structure and appear blurry. This objective is optimized using the schedule-free AdamW optimizer \citep{AdamW_Loshchilov_2017, schedule-free_Defazio_2024} for 100 epochs with a batch size of $32$, learning rate of $1.44 \times 10^{-4}$, $\epsilon= 10^{-8}$ and weight decay of $0.01$. The full VAE training workflow is discussed in Table \ref{tab:appx_vae_training}.

\subsubsection{Training of Denoising Network }

In our latent diffusion formulation, the diffusion process happens in latent space rather than directly on $\BB$, where the VAE is tasked to map the input magnetogram $\BB$ into a lower-dimensional latent representation $\zB \in \mathbb{R}^{h \times w \times 6}$ and from latent back to magnetograms $\hat{\BB} = \mathcal{D}(\zB) \approx \BB$.
With the VAE frozen, we train the denoising network jointly with the spatial rescaler $\mathcal{S}$ on paired \aia UV/EUV maps and \hinode vector magnetogram cutouts of size $128\arcsec \times 128\arcsec$ (where $128\arcsec$ balances data availability and cutout size).

Given input magnetogram $\BB$, we first encode it with the VAE encoder $\mathcal{E}$ to obtain its latent representation $\zB = \mathcal{E}(\BB)$. 
A timestep $t \in \{1, 2, \cdots, T \}$ is then randomly sampled and we add noise $\epsilonB$ following the forward process \autoref{eq:LDM_forward} to obtain $\tilde{\zB}_t$, the noisy magnetogram latent at timestep $t$. 
The conditioning input $\IB$ is projected by the spatial rescaler $\mathcal{S}$ to an $\mathbb{R}^{h \times w \times 11}$ tensor, matching the spatial dimensions of the magnetogram latent $\zB$.
We then broadcast the solar cycle indicator to latent resolution and concatenate it with the rescaled conditioning input, producing the full conditioning tensor $\cB \in \mathbb{R}^{h \times w \times 12}$.

$\tilde{\zB}_t$ and $\cB$ are then concatenated into a latent of shape $\mathbb{R}^{h \times w \times 18}$ and fed to the denoising network $f_\theta$, which predicts the velocity term $\hat{\vB}_t = f_\theta (\tilde{\zB}_t, \cB)$. The velocity term is then compared against the ground truth velocity $\vB_t$ to calculate loss. Here, we use $\ell_2$ as training objective following standard practice in stable diffusion~\citep{rombach_high-resolution_2022}. This objective is optimized with the schedule-free AdamW optimizer for $100$ epochs with batch size of $32$, learning rate of $3.2 \times 10^{-4}$, $\epsilon= 10^{-8}$ and no weight decay. The full denoising network training workflow is discussed in Table \ref{tab:appx_unet_training}.

\subsubsection{Inference Process}
With every network frozen, inference process begins with sampling pure Gaussian noise, denoted $\tilde{\zB}_T \sim \mathcal{N}(0,1)$, where $T$ is the total number of diffusion steps in the sampling schedule. The conditioning UV/EUV filtergrams are again projected to $\cB$. Then, for every timestep update determined by the DDIM diffusion scheduler $t \rightarrow t_{next}$, we repeatedly do: {[1]} Concatenate current noisy latent $\tilde{\zB}_t$ with $\cB$. {[2]} Feed the concatenated latent to denoising network $f_\theta$ and obtain predicted velocity at current timestep $\hat{\vB}_t$. {[3]} Use diffusion scheduler to remove noise from $\tilde{\zB}_t$ to obtain cleaner latent $\tilde{\zB}_{t_{next}}$. 

This reverse process is repeated until the initial noise is transformed into a clean latent $\hat{\zB}_0$. Finally, the decoder $\mathcal{D}$ maps this latent back to the predicted magnetogram $\hat{\BB} = \mathcal{D}(\hat{\zB}_0)$. The full inference workflow is discussed in Table \ref{tab:appx_full_inference_process}.

Since the inference process starts from random noise, each individual run produces a distinct sample from the learned conditional distribution $p(\BB \mid \IB)$, which we refer as \emph{realizations}. By generating $k$ realizations for the same input, we can assess the diversity of plausible magnetograms, estimate per-pixel uncertainty, and infer underlying correlations.

\begin{figure*}[tp]
    \centering
    \includegraphics[width=1\linewidth]{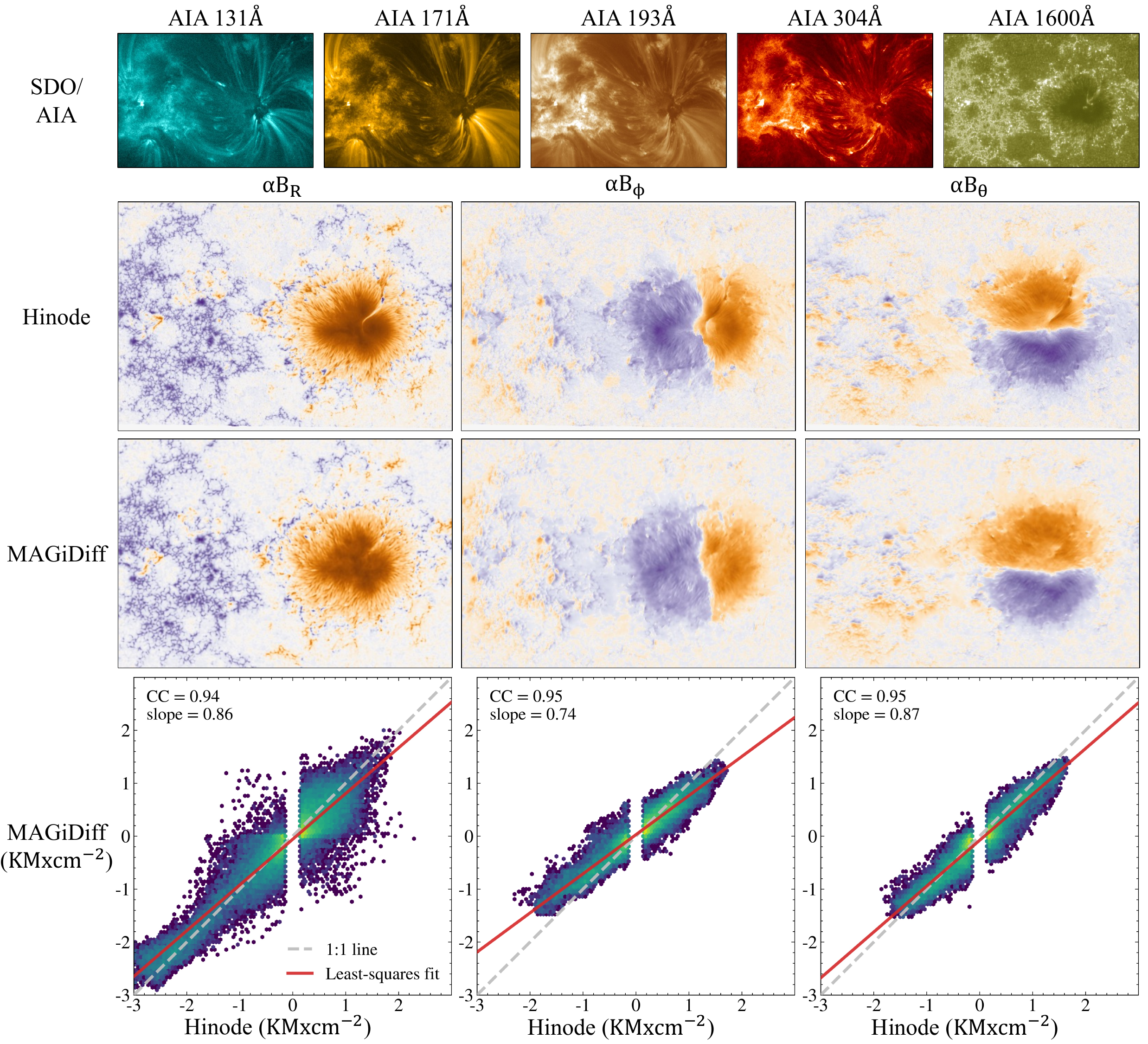}
    \caption{\textbf{A qualitative result of \modelname on the test set with \hinode as reference, with the corresponding pixel value distributions.} 
    Top row: selected input UV/EUV intensity images, left-to-right \aia 131\AA, 171\AA, 193\AA, 304\AA, 1600\AA.
    Middle rows: \hinode vector magnetogram and \modelname prediction, left-to-right \abr, \abp, \abt.
    Bottom row: hexbin density plot of \modelname predictions against \hinode ground truth.
    In the hexbin panels, pixels with ground truth absolute value below $150$ \gauss~ are omitted to prevent low-amplitude quiet-Sun pixels dominating the pixel population. The gray dashed line marks the $1{:}1$ relation and the red line marks the least-squares fit. Each panel also reports the Pearson correlation coefficient (CC) and slope of the best fit line.
    \modelname recovers the dominant magnetic structures and closely mimics \hinode. In faint plage and quiet-Sun regions, \modelname predictions are generally correct but can have reversed polarity and reduced detail. The hexbin panels confirm this agreement quantitatively, in which most pixel density concentrates near the $1{:}1$ relation.
    Example Date: 2016 April 11, 12:36 TAI. 
    Colormaps: -3000 \includegraphics[width=30pt,height=6pt]{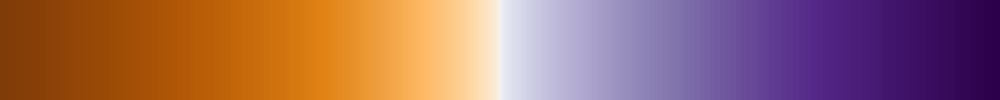} 3000 \gauss 
    using signed square root $x \mapsto \textrm{sign}(x) \sqrt{|x|}$ for contrast. Histogram legend: 1 \includegraphics[width=30pt,height=6pt]{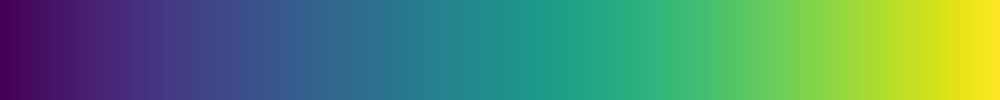} 1000 counts, shown logarithmically.
     }
    \label{fig:fig1_qualitative}
\end{figure*}

\section{Results} \label{sec:results}

We now describe the evaluation of \modelname on unseen data from five complementary perspectives. First, we assess \modelname on a per-pixel basis, treating it purely as a method that takes a set of UV/EUV filtergrams and produces a vector field. Next, we examine whether \modelname recovers the correct magnetic polarity. Having analyzed \modelname, we then turn to comparing it to a more standard regression approach to test its contribution. We further analyze the distribution learned by \modelname. Finally, we evaluate \modelname's generalization ability beyond the training setting, including full-disk observations and data from other instruments.

\subsection{Qualitative Results}
\label{sec:qualitative results}

\begin{figure*}[t]
    \centering
    \includegraphics[width=1\linewidth]{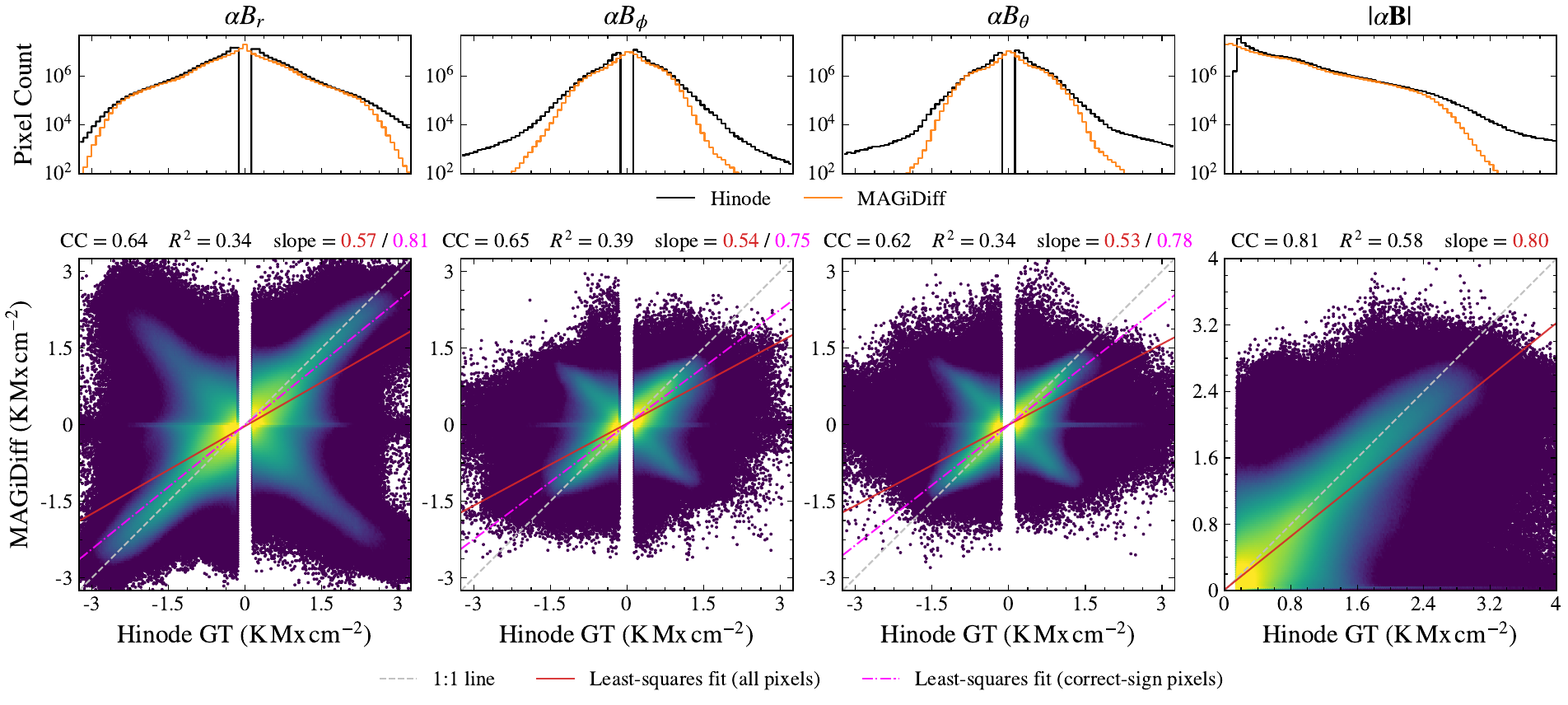}
    \caption{\textbf{Pixel-value histograms and hexbin density plots of \modelname predictions against \hinode vector magnetograms on the test set.}
    Each column corresponds to \abr, \abp, \abt, and $|\alpha \mathbf{B}|$. The top panels compare the distributions of \modelname predictions and the \hinode ground truth. The bottom panel shows the corresponding hexbin density plot, together with a gray dashed line marking the $1{:}1$ relation, a red line marking the least-squares fit on all pixels, and a purple dashed line marking the least-squares fit on pixels for which the predicted sign agrees with the ground truth. Each panel lists the Pearson correlation coefficient (CC) and the coefficient of determination ($R^2$), both computed over all plotted pixels, together with the slopes of the two fit lines. 
    Density concentrated near the $1{:}1$ line indicates agreement between \modelname prediction and ground truth. Deviations from this line are most apparent near the extremes of the ground-truth distribution, where \modelname tends to underestimate field strengths. 
    In \abr, \abp, \abt, a weaker concentration near the $y=-x$ relation indicates pixels for which \modelname predicts approximately the correct magnitude but opposite sign.
    The faint horizontal concentration near $y=0$ marks pixels for which \modelname predictions are close to zero despite having significant ground truth field strengths, usually due to misalignment or small magnetic fields without indications in UV/EUV. 
    To make the trends more apparent, pixels with ground truth absolute value below $150$~\gauss are omitted from all panels, since low-amplitude quiet-Sun pixels dominate the full pixel population. Histogram legend: $100$ \includegraphics[width=30pt,height=6pt]{colorbar_viridis.png} $100000$ count, shown logarithmically.}
    \label{fig:hexbin}
\end{figure*}

We first evaluate how well \modelname estimates \hinode-like vector magnetograms from previously unseen UV/EUV observations. As shown in \autoref{fig:fig1_qualitative}, \modelname recovers the locations, spatial distribution, and relative field strengths of the dominant magnetic structures observed by \hinode. The predictions preserve active magnetic structures while reproducing weaker surrounding fields and fine-scale variations in quiet-Sun regions.

To examine the agreement beyond the overall visual appearance, \autoref{fig:fig1_qualitative} also presents hexbin plots comparing the predicted and \hinode field values. The predictions show strong agreement with \hinode over a broad range of field strengths, with a tendency to underestimate the strongest field values.

\subsection{Test Set Distributional Agreement}
\label{sec:scatter}

We next examine the joint distributions of the pixel values predicted by \modelname and the corresponding values in \hinode magnetograms across the full test set, as shown in \autoref{fig:hexbin}. Overall, \modelname reproduces the broad distributions of the vector field components, with dominant pixel density concentrated along the $1{:}1$ relation. Deviations are mainly seen at the highest field strengths, where \modelname tends to underestimate the most extreme values. 

For \abr, \abp, and \abt, we additionally observe two weaker branches concentrated near the $y=-x$ and $y=0$ relations. 
Pixels near the $y=-x$ relation have approximately the correct magnitude but the opposite sign. Visual inspection of the contributing samples suggests two main types of polarity errors, illustrated by representative examples in \autoref{fig:fig_hexbin_incorrect_exp}. 
First, in complex active regions containing multiple magnetic structures with different polarities, \modelname may assign the wrong polarity to individual structures. 
Second, \modelname may recover the large-scale polarity configuration correctly while assigning incorrect polarity to localized plage regions. Despite the polarity reversal, the predicted magnetic structures remain spatially plausible and broadly agree with the ground truth. These errors reflect the difficulty of disambiguating polarity, especially in regions where the UV/EUV observations provide no clear polarity cue.

The $y=0$ branch corresponds to pixels for which \modelname predicts a value near zero despite a nonzero ground truth value. This branch largely arises from two sources. First, \modelname fails to recover some relatively weak-field structures, causing the predicted values to collapse toward zero. 
Second, the difficulty of inferring pixel-accurate magnetic structures can produce small spatial misalignments, causing pixels near magnetic structure boundary to have nonzero ground truth but near-zero predictions.
These two error modes reflect the difficulty of using limited and indirect information provided by UV/EUV filtergrams to recover weak magnetic fields and sharp magnetic structure boundaries.

\subsection{Polarity Agreement}
\label{sec:polarity_accuracy}

\begin{deluxetable*}{lcccccccccc}[t]
\tablewidth{0pt}
\tabcolsep=5.5pt
\tablecaption{\textbf{The polarity prior improves \modelname's overall polarity agreement. It is most effective where a clear leading--following polarity pair exists: bipolar $\beta$ regions gain the most, while unipolar $\alpha$ regions are frequently wrong with or without the prior and the most complex $\beta\gamma\delta$ regions remain difficult.} We evaluate \modelname with and without the polarity prior on the test set, using one realization per input for each variant. The overall columns report the fraction of predictions with $\tau$ below each threshold, where $\tau$ is defined in \autoref{eqn:polarity_ratio}: $\tau < 1.0$ indicates that the predicted polarity agrees with the ground truth better than its sign-flipped counterpart, while $\tau < 0.5$ gives a stricter measure of a confident polarity agreement. The remaining columns report the wrong-polarity rate ($\tau \geq 1$, lower is better) among valid test samples of each Mount Wilson active-region class, where each sample is labeled by the most complex NOAA active region within its valid ground-truth footprint and QS/Plage marks frames with no spotted region. The polarity prior column indicates whether latitude and solar cycle inputs are included during training. \label{tab:polarity_by_ar_class}}
\tablehead{\colhead{Model} & \colhead{Polarity Prior} & \multicolumn{2}{c}{Overall polarity $\uparrow$} & \multicolumn{7}{c}{Wrong-polarity rate by AR class $\downarrow$}\\\colhead{} & \colhead{} & \colhead{Pass ($\tau\!<\!1$)} & \colhead{Conf.\ ($\tau\!<\!0.5$)} & \colhead{QS/Plage} & \colhead{$\alpha$} & \colhead{$\beta$} & \colhead{$\beta\delta$} & \colhead{$\beta\gamma$} & \colhead{$\beta\gamma\delta$} & \colhead{All}}
\startdata
    Samples &  &  &  & 1061 & 455 & 2490 & 112 & 1099 & 1077 & 6294 \\
    Distinct NOAA ARs &  &  &  & -- & 19 & 61 & 4 & 30 & 18 & 93 \\
    \hline
    MAGiDiff $\pm\BB$ & $\times$ & 48.7 & 41.5 & 51.0 & 47.7 & 50.9 & 68.8 & 52.3 & 51.4 & 51.3 \\
    MAGiDiff No Polarity Prior & $\times$ & 86.7 & 76.6 & 15.3 & 20.0 & 12.7 & 8.9 & 10.0 & 13.9 & 13.3 \\
    MAGiDiff & \checkmark & \textbf{92.7} & \textbf{84.8} & \textbf{9.1} & \textbf{8.4} & \textbf{4.2} & \textbf{0.0} & \textbf{8.6} & \textbf{12.0} & \textbf{7.3} \\
\enddata
\end{deluxetable*}

Beyond field strength and spatial structure, we also evaluate a physically important property: whether \modelname recovers the correct magnetic polarity.
Throughout the results section, we use \emph{polarity} to refer to the sample-level three-dimensional vector orientation of the vector magnetogram. Here, $\hat{\BB}$ denotes the predicted magnetic field, and $-\hat{\BB}$ denotes the same field with the signs of all three components (\abr, \abp, \abt) simultaneously flipped. 

UV/EUV filtergram intensities are sign-invariant, and therefore are not sufficient for determining polarity. Hale's law ~\citep{Hale1925} nevertheless could be used to resolve this ambiguity: for bipolar active regions, the expected leading polarity depends on the hemisphere and reverses between successive solar cycles. Thus, UV/EUV filtergrams, heliographic latitude, and solar cycle information may together allow the \modelname to infer the correct polarity.
Because the \aia filtergrams are physically corrected for instrumental degradation, solar cycle parity should not, in principle, be inferable from the filtergrams alone. This reasoning motivates us to provide \modelname with an explicit \emph{polarity prior}, consisting of a per-pixel heliographic latitude map and a binary solar cycle indicator.

We quantify polarity agreement using a score, $\tau$, defined as the ratio between the summed square error for the predicted field $\hat{\BB}$ to that of its sign-flipped counterpart $-\hat{\BB}$, both measured against the ground truth $\BB$:
\begin{equation}
\label{eqn:polarity_ratio}
\tau(\hat{\BB}, \BB) \coloneqq
\frac{\lVert \hat{\BB} - \BB \rVert_2^2}{\lVert -\hat{\BB} - \BB \rVert_2^2 + \delta},
\end{equation}
where $\delta$ is a small constant included for numerical stability.
We compute $\tau$ after downsampling the prediction and ground truth by a factor of $4$ to reduce sensitivity to pixel-scale misalignment, and restrict the evaluation to pixels with $|\alpha \BB| > 150$~\gauss. A value of $\tau < 1$ indicates that the predicted polarity matches the ground truth better than the sign-flipped version; smaller values indicate greater confidence.

\subsubsection{Effect of the Polarity Prior}

\begin{figure*}[t]
    \centering
    \includegraphics[width=1\linewidth]{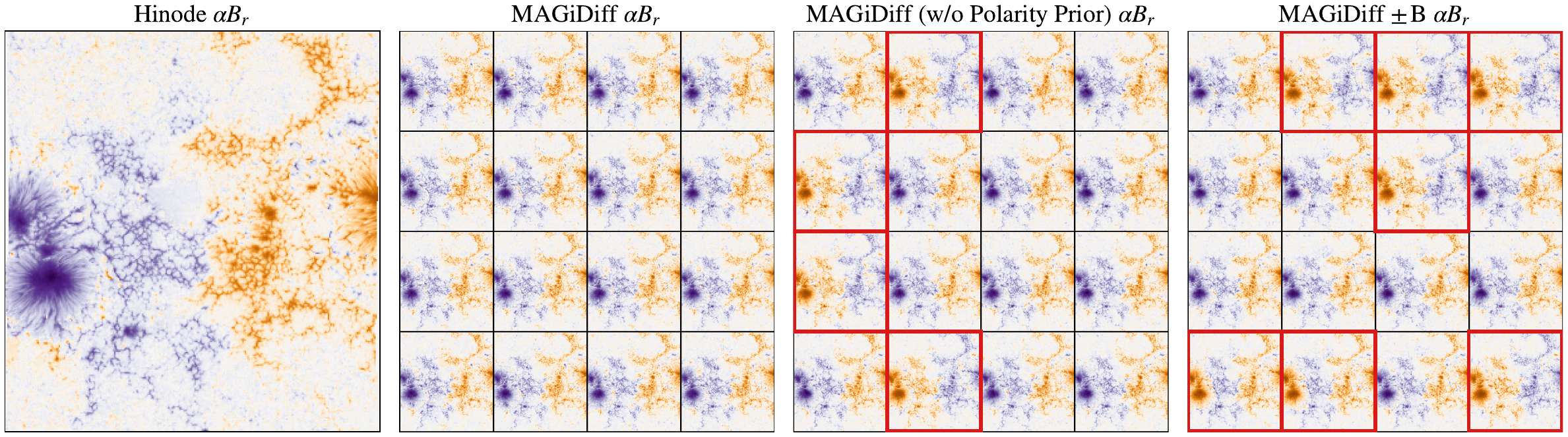}
    \caption{\textbf{ Representative qualitative comparison of \modelname, \modelname without polarity prior, and \modelname $\pm \BB$.} 
    From left to right, the figure shows the \hinode \abr reference, followed by $16$ realizations generated by \modelname, \modelname without polarity prior, and \modelname $\pm \BB$, respectively. Realizations with incorrect polarity are outlined in red.
    Example date: 2016 September 5, 14:48 TAI. 
    Colormaps: -3000 \includegraphics[width=30pt,height=6pt]{color_PuOrSqrt.png} 3000 \gauss following \autoref{fig:fig1_qualitative}.
     }
    \label{fig:fig_polarity_grid}
\end{figure*}

To understand how \modelname resolves polarity ambiguity, we evaluate three training formulations.
\begin{itemize}
    \item 
(\textbf{\modelname $\pm \BB$}) trains \modelname to produce {\it both} $\BB$ and $-\BB$ as equally valid training targets, since both orientations yield the same UV/EUV intensities. Because the training objective has no preference for either orientation, the model should randomly select between the two. 
\item (\textbf{\modelname No Polarity Prior}) uses the real \hinode magnetogram as the target, without introducing any $\BB$ vs $-\BB$ ambiguity into the training objective. Theoretically, UV/EUV inputs contain no direct information that is correlated with photospheric magnetic polarity, so the model should perform close to chance.
\item (\textbf{\modelname}) is \modelname as described in Section~\ref{sec:method}, including the polarity prior. The polarity prior provides \modelname with information that is needed in order to use Hale's law to estimate the polarity.
\end{itemize}

We report results in \autoref{tab:polarity_by_ar_class}, comparing all three methods. We additionally provide a qualitative comparison on a representative sample in \autoref{fig:fig_polarity_grid}. As expected, \textbf{\modelname $\pm \BB$} produces near-chance performance. Similarly, since \textbf{\modelname} has access to information about solar cycle and latitude that permits the use of Hale's law, it is able to obtain the polarity quite accurately, at 92.7\% accuracy. This demonstrates that the network can take advantage of this signal and assign polarities. However, surprisingly \textbf{\modelname No Polarity Prior} performs {\it far} above chance, at 86.7\%. This unexpectedly strong performance motivated us to investigate what provided the missing polarity information.

Our suspicion was that the model was exploiting Hale's law by implicitly obtaining the information needed to use the law to assign polarity, namely latitude and solar cycle number. We therefore conducted auxiliary experiments to determine if these could be determined from UV/EUV filtergrams. We trained basic ResNet-50~\citep{he2015deepresiduallearningimage} models for northern-vs-southern hemisphere and solar cycle 24-vs-25, and evaluated following the split in Section~\ref{sec:data}.

Northern-vs-southern Hemisphere was relatively easy for the classifier, with 94.7\% accuracy and an AUROC of 0.998. Its performance degrades only near the equator, from $97.8\%$ for crops beyond $\pm 25^\circ$ in heliographic latitude to $73.7\%$ for crops within $\pm 10^\circ$. These results indicate that the classifier exploits geometric cues such as foreshortening and center to limb intensity variations to infer hemisphere.

More surprisingly, the solar cycle classifier reaches $98.5\%$ accuracy and an AUROC of $0.996$. This result is unexpected because UV/EUV filtergrams do not encode time, and the \aia degradation correction is intended to remove long-term instrumental changes. Thus, the solar cycle should not be detectable in, for instance, the average intensity. However, since the instrument throughput has {\it decreased}, the signal to noise ratio has changed. As a simplified worked example, consider a light source that at mission start would produce a $100$ \dnpersec count on the detector. Assuming a Poisson model, this measurement would have a standard deviation of $10$ \dnpersec. After a $5\times$ degradation in filter performance over the mission, the same light would yield $20$ \dnpersec with a standard deviation of $\sqrt{20}$. The standard degradation correction increases the signal $5\times$, recovering the original $100$ \dnpersec, but with a final noise strength of $5 \sqrt{20} = 22.4$ \dnpersec. The change in noise strength would leave a signature in the images, which would be visible in differences across the pixels, especially in channels with strong degradation and low count rates.

Channel ablations provide evidence that the network may be doing this. We retested the network while replacing each channel with its dataset mean in order to test dependence. 304\AAs has a relatively low count rate and has had a nearly $10\times$ drop in measured strength over the mission. Performance in classifying the solar cycle drops precipitously to an accuracy of 50\% when 304\AAs is replaced. On the other hand, 1700\AAs has remained steady and replacing 1700\AAs with the dataset mean has virtually no effect on performance.

Together, these suggest that there are real signals that can be exploited in the passbands after correction, allowing neural networks to infer solar cycle polarity.
However, inferring these information indirectly from UV/EUV filtergrams, particularly solar cycle parity from instrument-specific degradation, is less robust than providing it as input directly. We therefore supply \modelname with per-pixel heliographic latitude map and a binary solar cycle indicator as \emph{polarity prior}. And indeed adding this explicit prior further increases overall polarity accuracy.

\subsubsection{Classification by AR type}

To identify the magnetic configurations in which \modelname reliably recovers polarity and those in which it fails, we divide the test samples according to Mount Wilson active region class~\citep{Hale_AR_class_1919, Jaeggli_AR_class_2016} following the method described in~\cite{ARCAFF_ar_classification_dataset}. Each sample is assigned the class of the most complex NOAA active region~\citep{Jaeggli_AR_class_2016} within its valid ground truth footprint. We further validate candidate associations against the observations using sunspot darkness in \hmi continuum map, plage brightness in co-aligned \aia 1600\AAs filtergram, and magnetogram structure in the ground truth \hinode vector magnetogram. Samples with no corresponding active region are grouped as quiet sun (QS) or plage. We also exclude near-limb samples for which at least half of the valid pixels have $\mu<0.35$, where $\mu$ is the cosine of the viewing angle. These samples are strongly affected by foreshortening and yield less reliable measurement.

\autoref{tab:polarity_by_ar_class} reports the wrong polarity rate for each class, defined as the fraction of samples with $\tau \geq 1$. As expected, \modelname $\pm \BB$ performs near chance and \modelname with polarity prior does the best in all cases. We therefore focus on how performance varies across active region classes.

The clearest improvement occurs for the simple bipolar $\beta$ regions, to which Hale's law is most directly applicable. The polarity prior also improves polarity recovery for the more complex bipolar $\beta\delta$ and $\beta\gamma$ regions. However, the $\beta\delta$ result should be interpreted cautiously because this class contains only $112$ samples from $4$ distinct active regions. 
For the most complex $\beta\gamma\delta$ regions, the improvement is small, as their intermingled multipolar structure usually lacks a well-defined leading-following organization, limiting the applicability of Hale's law. 
Surprisingly, \modelname also predicts the polarity of unipolar $\alpha$ regions with high accuracy. A plausible explanation is that many $\alpha$ regions are evolved remnants of bipolar regions. The UV/EUV morphology may therefore indicate whether the magnetic structure corresponds to the leading or following part of the original bipolar structure, allowing the usage of Hale's law.

\begin{figure*}[tp]
    \centering
    
    \includegraphics[width=1\linewidth]{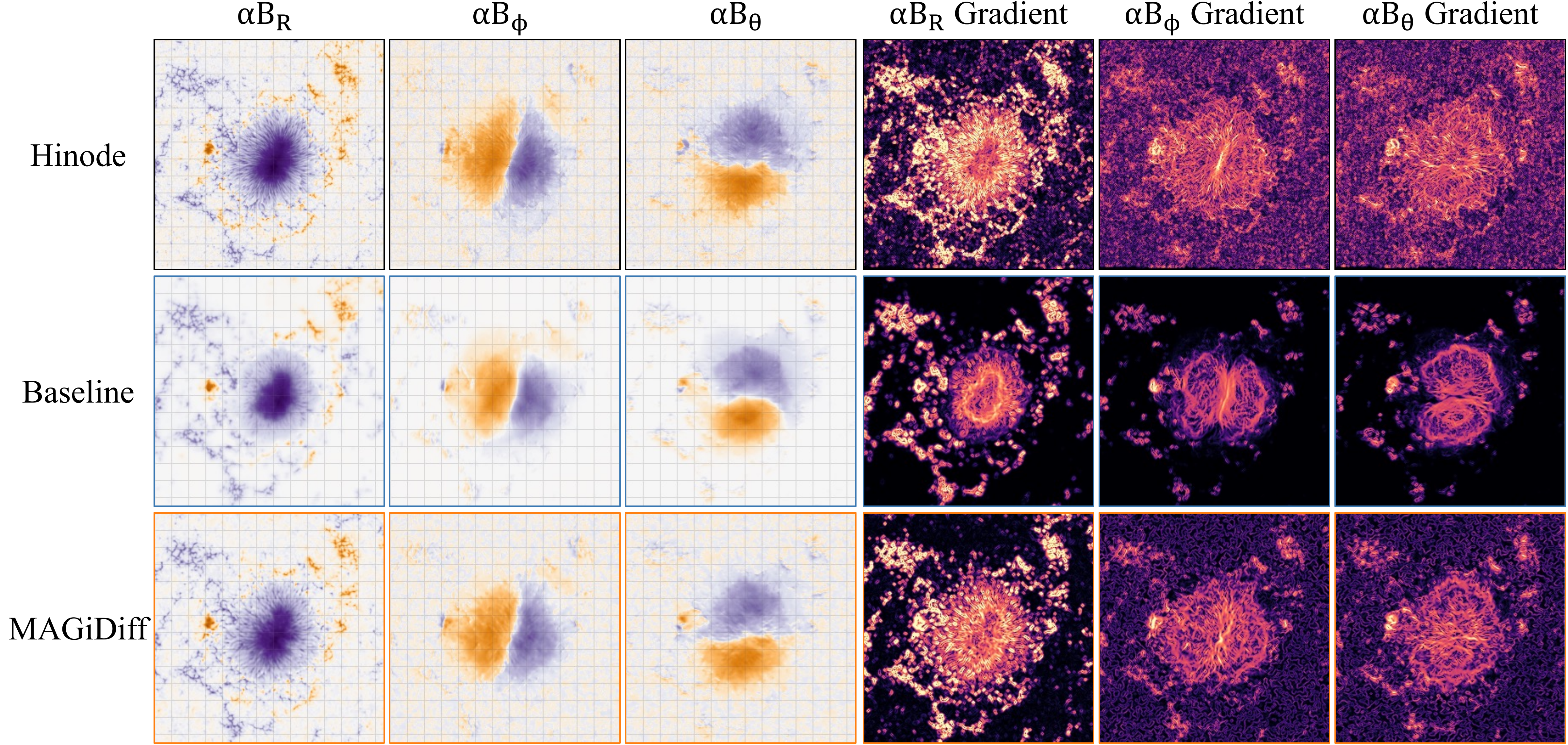}
    \caption{\textbf{Qualitative comparison of \modelname and Regression U-Net baseline, with the corresponding gradient magnitude maps.} 
    Example date: 2016 June 15, 03:36 TAI. Both \modelname and the regression baseline accurately predict the large-scale magnetic field structure. However, \modelname additionally reproduces the realistic quiet-Sun texture as seen in \hinode, whereas the regression baseline collapses the quiet Sun to an nonphysically smooth background.
    The gradient maps show that this smoothing is systematic, where the baseline underestimates the gradient magnitude in every component.
    Magnetogram colormap: -3000 \includegraphics[width=30pt,height=6pt]{color_PuOrSqrt.png} 3000 \gauss following \autoref{fig:fig1_qualitative}. Gradient magnitude colormap: $0$ \includegraphics[width=30pt,height=6pt]{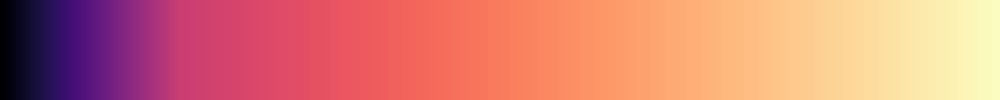} $250$ \gpx. 
     }
    \label{fig:baseline}

    \vspace{0.1cm}
    \includegraphics[width=1\linewidth]{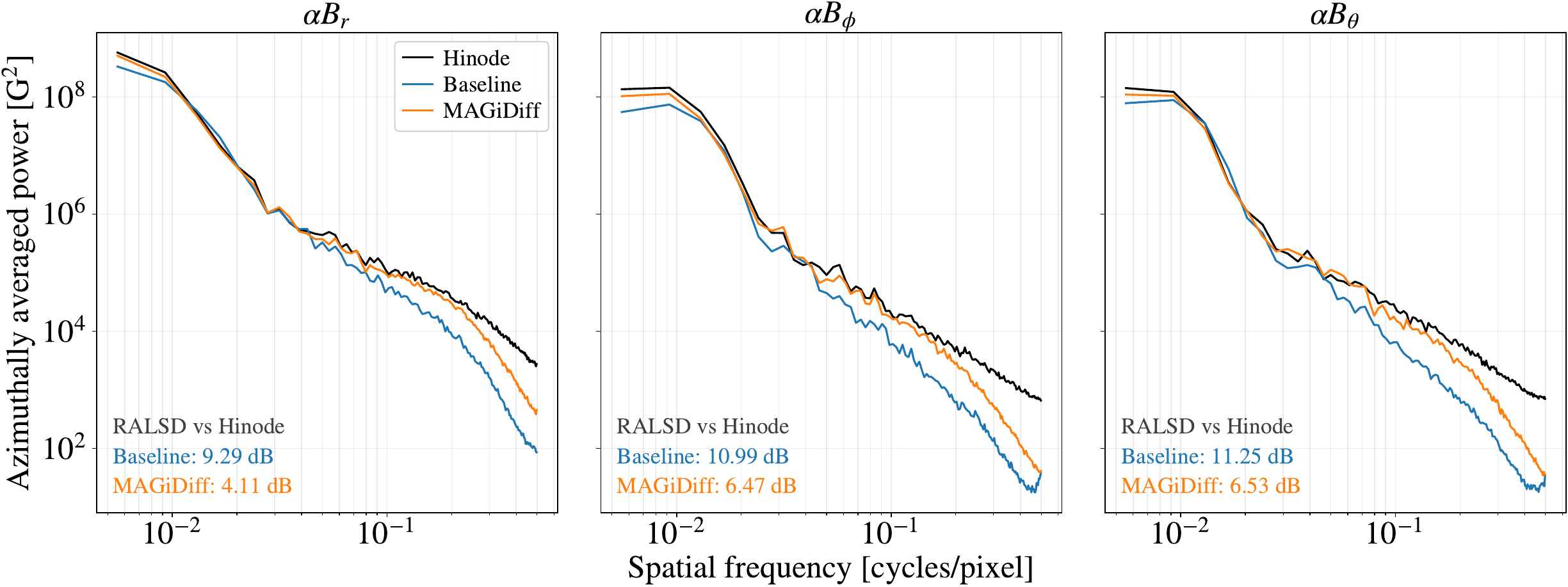}
     \caption{\textbf{Power spectrum of \modelname, regression U-Net baseline, and \hinode for sample in \autoref{fig:baseline}.} 
     All three spectra agree on large scale features, confirming that both models recover the large-scale field. However, they separate at fine scale details, where baseline's spectra fall from \hinode while \modelname tracks it more closely. The spectra quantify what the gradient maps show qualitatively: the baseline smooths out fine scale details.
     }
    \label{fig:baseline_power}
    
\end{figure*}

\subsection{Comparison with Regression Baseline}
\label{sec:baseline comparison}

We next ask whether this reconstruction requires a stochastic generative model or can instead be achieved with a standard deterministic approach. Because the input filtergrams and target vector magnetograms are spatially aligned, a natural starting point is to formulate the task as image-to-image regression.   To this end, we train a regression U-Net using exactly the same training data as \modelname with an $\ell_2$ objective and compare its predictions with those of \modelname. The regression U-Net produces a single deterministic magnetogram for each input, whereas \modelname can generate multiple stochastic realizations by sampling from the learned conditional distribution. This distinction is particularly relevant because magnetic structure is only indirectly encoded in the UV/EUV filtergrams, leaving local magnetic field weakly constrained, particularly in weak-field regions. Multiple fine-scale magnetic structure configurations may therefore be consistent with the same input. Under an $\ell_2$ objective, the regression model tends to predict the average across these possibilities and therefore loses fine-scale detail. Comparing the two approaches therefore allows us to assess how modeling a conditional distribution, rather than producing a single regression estimate, affects reconstruction accuracy and structural realism.

\subsubsection{Qualitative Comparison}

We begin with qualitative inspection in \autoref{fig:baseline}, where we show \modelname and the regression U-Net baseline side by side. \modelname produces sharper structures, finer details, and quiet-Sun regions that more closely resemble the \hinode vector magnetograms. The regression baseline recovers the dominant active structure, but tends to produce smoother and more spatially averaged fields where compact features are weakened, sharp boundaries are softened, and weak-field regions appear overly quiet compared with the \hinode observations. 

The corresponding power spectra in \autoref{fig:baseline_power} provide complementary frequency-domain evidence. Both models reproduce the large-scale power measured by \hinode, but the baseline increasingly loses power at finer spatial scale, whereas \modelname remains closer to the \hinode spectrum.

\subsubsection{Metrics}

\begin{deluxetable*}{llc@{\;}c@{\;}cc@{\;}c@{\;}cc@{\;}c@{\;}cc@{\;}c@{\;}c}
\tablewidth{0pt}
\tabletypesize{\footnotesize}
\tabcolsep=11pt
\tablecaption{\textbf{\modelname matches the regression baseline in per-pixel accuracy while substantially better reproducing the pixel-value distribution, sharpness, and power spectrum of the ground truth.}
We compare \modelname with a regression U-Net baseline on the test set, using one prediction per input: a single sampled realization for \modelname and the deterministic output of the baseline; each cell reports Baseline\,$|$\,\modelname, with the better value in \textbf{bold}.
For each field component, we report the mean absolute error (MAE, lower is better) and the percentage of pixels with error below $300$~\gauss~(\%$<300$, higher is better), computed over strong-field pixels with $|\alpha \mathbf{B}| > 1000$~\gauss.
The remaining metrics measure how well predictions reproduce the structure of the ground truth over all valid pixels: $W_1$ reports the Wasserstein distance between the predicted and observed pixel-value distributions (lower is better); the $\nabla$-ratio compares the total gradient magnitude of the prediction to that of the ground truth, where values near 1 indicate matched sharpness and values below 1 indicate over-smoothing; $\nabla$-MAE reports the mean absolute error of the gradient magnitude (lower is better); and RALSD reports the radially averaged log-spectral distance between the predicted and observed power spectra (lower is better), computed on $256\times256$ crops for comparability. All metrics are computed after aligning the global polarity of each prediction with the \hinode ground truth.
\label{tab:summary}}
\tablehead{\multicolumn{1}{l}{Measures} & \multicolumn{1}{l}{Metric} & \multicolumn{3}{c}{\abr} & \multicolumn{3}{c}{\abp} & \multicolumn{3}{c}{\abt} & \multicolumn{3}{c}{$|\alpha\BB|$}}
\startdata
    \multicolumn{1}{l}{Strong-field accuracy} & \multicolumn{1}{l}{MAE [\gauss] $\downarrow$} & 612.5 & $|$ & \textbf{575.5} & \textbf{304.1} & $|$ & 304.3 & 296.4 & $|$ & \textbf{294.6} & 428.6 & $|$ & \textbf{372.8} \\
    \multicolumn{1}{l}{Strong-field accuracy} & \multicolumn{1}{l}{\%$<$300 $\uparrow$} & 46.3 & $|$ & \textbf{49.1} & 69.1 & $|$ & \textbf{69.8} & 70.6 & $|$ & \textbf{71.3} & 49.5 & $|$ & \textbf{57.0} \\
    \multicolumn{1}{l}{Value-distribution match} & \multicolumn{1}{l}{$W_1$ [\gauss] $\downarrow$} & 36.6 & $|$ & \textbf{14.9} & 30.7 & $|$ & \textbf{10.5} & 32.4 & $|$ & \textbf{10.9} & 64.8 & $|$ & \textbf{24.7} \\
    \multicolumn{1}{l}{Global sharpness match} & \multicolumn{1}{l}{$\nabla$-ratio $\rightarrow\!1$} & 0.39 & $|$ & \textbf{0.73} & 0.22 & $|$ & \textbf{0.62} & 0.21 & $|$ & \textbf{0.60} & 0.44 & $|$ & \textbf{0.71} \\
    \multicolumn{1}{l}{Local sharpness accuracy} & \multicolumn{1}{l}{$\nabla$-MAE [\gauss/px] $\downarrow$} & 41.84 & $|$ & \textbf{38.05} & 26.92 & $|$ & \textbf{20.93} & 27.35 & $|$ & \textbf{20.96} & 35.87 & $|$ & \textbf{34.72} \\
    \multicolumn{1}{l}{Power-spectrum match} & \multicolumn{1}{l}{RALSD [dB] $\downarrow$} & 8.56 & $|$ & \textbf{4.25} & 11.32 & $|$ & \textbf{6.00} & 11.55 & $|$ & \textbf{5.96} & 7.98 & $|$ & \textbf{4.04} \\
\enddata
\end{deluxetable*}

We next quantify these differences using metrics that measure both per-pixel accuracy and structural fidelity.
Before computing any of the metrics below, we align the polarity of each prediction with the corresponding \hinode observation by comparing $\hat{\BB}$ with $-\hat{\BB}$ and retaining the one with better fit. This is applied to both models to prevent a small number of predictions with flipped polarity from dominating the metrics. With this, the comparison therefore focuses on the accuracy of the recovered field strengths and spatial structure rather than the polarity recovery, which we evaluate separately in earlier section.
For pixel-wise accuracy, we follow \cite{higgins_synthia_2022, wang_supersynthia_2024} and report:
\begin{enumerate}
    \item {\bf the mean absolute error (MAE)}, or the average prediction error 
    \item {\bf the percentage of pixels with absolute error below a fixed threshold ($\% < t$; \cite{Scharstein02})}, with $t=300$ \gauss~ here. This metric measures the fraction of ''good'' pixels whose errors fall below a specified tolerance.
\end{enumerate}    
Following~\cite{wang_supersynthia_2024}, we evaluate them on strong-field pixels with $|\alpha \mathbf{B}| > 1000$~\gauss to avoid having metrics that are dominated by the far more numerous quiet-sun pixels.

These metrics only report pixel-to-pixel accuracy, and so to provide a more holistic picture, we report four additional metrics that capture distributional and structural fidelity.
\begin{enumerate}
    \setcounter{enumi}{2}
    \item \textbf{1-Wasserstein distance ($W_1$)}. Also known as the Earth Mover's Distance (EMD; \cite{Rubner2000TheEarthMoving}) in computer vision, $W_1$ \citep{W1origin_Kantorovitch_1958} measures the minimum transformation cost of transporting probability mass between two distributions. For each field component, we construct empirical one-dimensional distributions from valid pixel values in the prediction and ground truth, and compute the transport cost required to transform one distribution into the other. Thus, $W_1$ measures agreement between the predicted and ground truth distributions without considering the spatial locations of individual pixels, where lower values indicate better agreement.

    \item \textbf{$\nabla$-ratio}. Spatial gradients characterize the strength and location of local magnetic field variations. Preserving these quantities is important not only for reproducing magnetic structures, but also for downstream calculations of physical quantities based on field derivatives, for example electric current density. $\nabla$-ratio is the ratio of the total gradient magnitude in the prediction to that in the ground truth. It is measured by aggregating gradient magnitude over the entire image, providing a global measure of whether the model preserves the overall level of spatial variation. Values near unity indicate matched overall sharpness, whereas values below unity indicate over-smoothing, and values above unity indicate excessive spatial variation, which may result from overly sharp structures or noise.

    \item  \textbf{$\nabla$-MAE}. Because $\nabla$-ratio compares only the aggregated gradient magnitudes, predictions can have similar values even when the variations occur at incorrect locations. We therefore also report $\nabla$-MAE, the mean absolute error on per-pixel gradient magnitudes. This metric complements the global $\nabla$-ratio by penalizing discrepancies in the locations and strengths of local spatial variations. 

    \item \textbf{Radially averaged logarithmic spectral distance (RALSD)}. RALSD \citep{RALSD_Harris_2022} compares the predicted and ground truth power spectra, with lower values indicating better agreement. To compute this, we first apply a two-dimensional Fourier transform to each sample and calculate the power spectrum as the squared magnitude of the Fourier coefficients. We then azimuthally average the power over concentric rings in frequency space, producing a one-dimensional radial power spectrum indexed by spatial frequency magnitude. RALSD is then computed as the root mean square error, across radial frequency bins, between the predicted and ground truth power spectra after conversion to decibels. For comparability across samples, we compute RALSD on a $128\arcsec \times 128\arcsec$ cutout centered within the largest fully valid region of each test sample. RALSD therefore measures whether the prediction distributes the correct amount of power across spatial scales, from large-scale magnetic structure to fine-scale details. 
\end{enumerate}

\begin{figure*}[tp]
    \centering
    \includegraphics[width=0.9\linewidth]{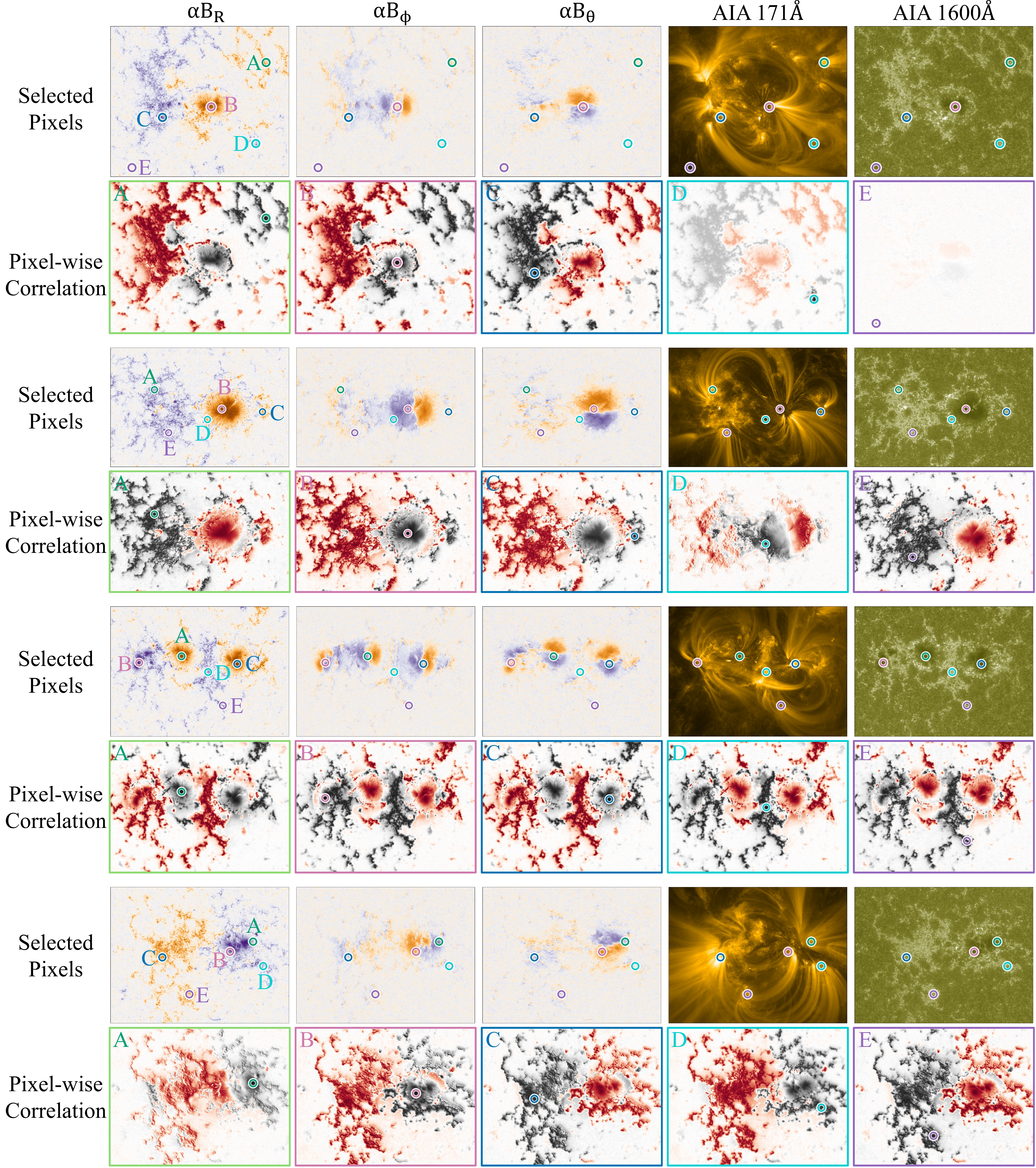}
    \caption{\textbf{Pixel-wise correlation map on four test examples from 2016.}
   Dates, top to bottom: Jan~26, 23:12 TAI; Apr 14, 02:48 TAI; Jul 19, 16:48 TAI; Oct~08, 09:12 TAI. 
    For each panel:
    \textbf{Top Row:}  Left to right: $\alpha B_R$, $\alpha B_\phi$, $\alpha B_\theta$, \aia $171$\AA, $1600$ \AA.
    \textbf{Bottom Row:} Pixel-wise correlation of 5 selected pixels (labeled \textbf{A-E}) with respect to all other pixels.
    As an example, we discuss results in the first panel with clear correlation clue.
    The sunspot pixel (\textbf{B}) accurately correlates with the sunspot region and its connected plage.
    The closed-loop footpoint (\textbf{C}) shows correlations that span the entire loop and show the correct opposite polarity at the conjugate footpoint. 
    The quiet region (\textbf{E}) has near-zero correlations as expected. 
    For reference, we also show ground truth \hinode $\alpha B_R$ in \autoref{fig:appx_correlation_gt}.
    Colormaps: -3000 \includegraphics[width=30pt,height=6pt]{color_PuOrSqrt.png} 3000 \gauss for $\alpha B_R$, $\alpha B_\phi$, $\alpha B_\theta$ following \autoref{fig:fig1_qualitative}; -1 \includegraphics[width=30pt,height=6pt]{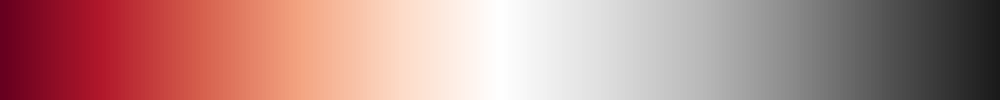} 1 for correlation maps. ($1$: correlated; $0$: uncorrelated;  $-1$: anti-correlated). 
   }
    \label{fig:appx_correlation}
\end{figure*}

\subsubsection{Quantitative Results}

Table~\ref{tab:summary} reports the quantitative comparison between the regression baseline and \modelname. For a fair comparison with the deterministic regression baseline, we evaluate \modelname using a single stochastic realization per input.
On strong-field pixels, \modelname outperforms the regression baseline on almost all field components. It achieves lower MAE and higher $\%<300$ percentage for  \abr, \abt, and $|\alpha\mathbf{B}|$, while remaining comparable for \abp. Thus, even in the single realization setting, \modelname matches and often exceeds the regression baseline's performance, measured on a per-pixel basis.
The distributional and structural metrics show larger differences.
\modelname achieves much lower $W_1$ and RALSD, indicating better agreement with \hinode in both pixel-value distribution and the allocation of power across spatial scales. The $\nabla$-ratio shows that \modelname preserves the total gradient magnitude present in \hinode more closely. The lower $\nabla$-MAE achieved by \modelname complements this global comparison by showing better agreement in the strength and location of local spatial variations. Together, these metrics confirm that \modelname more faithfully reproduces structural and distributional properties of \hinode magnetograms.

\subsection{Statistical Characteristics of the Predictive Distribution}

\begin{figure*}[t]
    \centering
    \includegraphics[width=1\linewidth]{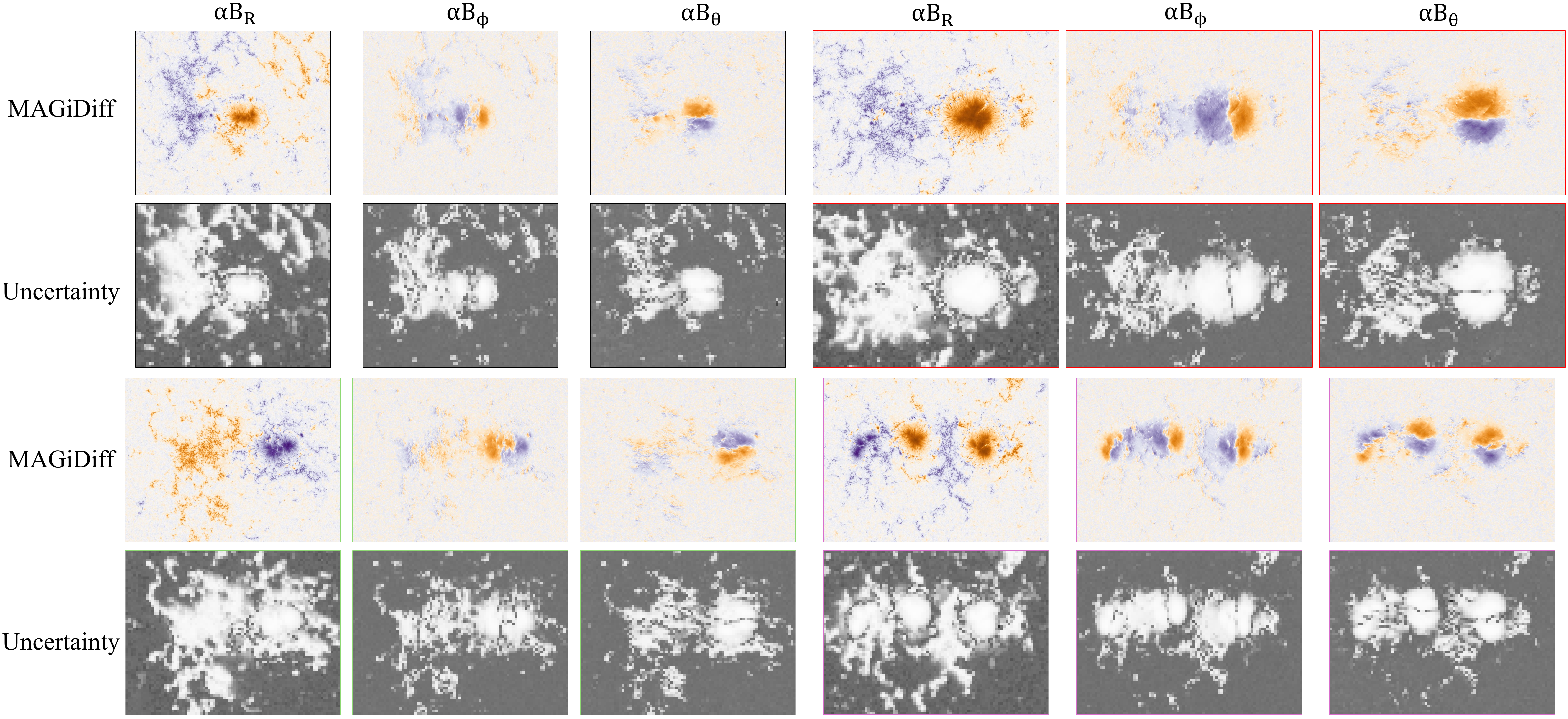}
    \caption{\textbf{Uncertainty map for four cutouts in \autoref{fig:appx_correlation}.} For each sample in \autoref{fig:appx_correlation}, we present \modelname's uncertainty estimates, defined as the per-pixel standard deviation divided by the per-pixel mean magnitude (a unitless measure).
    Example date: 2016 January 26, 23:12 TAI; 2016 April 14, 02:48 TAI; 2016 October 8, 09:12 TAI; 2016 July 19, 16:48 TAI. 
    For each panel:
    \textbf{Top Row:}  Left to right: $\alpha B_R$, $\alpha B_\phi$, $\alpha B_\theta$.
    \textbf{Bottom Row:} Left to right: Uncertainty map for $\alpha B_R$, $\alpha B_\phi$, $\alpha B_\theta$.
    Colormaps: -3000 \includegraphics[width=30pt,height=6pt]{color_PuOrSqrt.png} 3000 \gauss for $\alpha B_R$, $\alpha B_\phi$, $\alpha B_\theta$ following \autoref{fig:fig1_qualitative}; 0 \includegraphics[width=30pt,height=6pt]{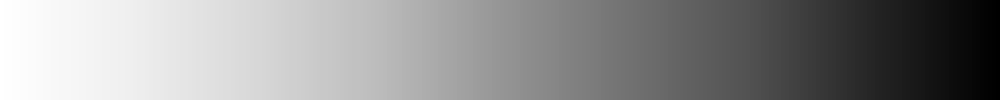} 2 (unitless) for uncertainty maps. 
   }
    \label{fig:appx_uncertainty}
\end{figure*}

Beyond evaluating individual predictions, we examine whether the predictive distribution learned by \modelname respects physical relationships among magnetic structures. 
In particular, a useful distribution should encode relationships among pixels: pixels belonging to the same magnetic structure should tend to share the same polarity across realizations, pixels rooted at opposite ends of a closed magnetic loop should exhibit opposite polarity, while unrelated quiet-Sun pixels should show little polarity correlation. The probabilistic formulation of \modelname allows us to probe these relationships directly by sampling ensembles of realizations.

Recall that we aimed to learn a model that could draw samples from $p(\BB | \IB)$. In practice, the model's inferred distribution is not necessarily equal to the actual conditional distribution \citep{wu_2024_Katie}, and so let us denote the model's actual distribution $\hat{p}(\BB | \IB)$. 
To quantify the statistical relationship between the two image locations $i$ and $j$, we compute the mean cosine similarity between their magnetic field vectors across realizations: 
\begin{equation}
\label{eqn:similarity}
C_{i,j}(\IB) \coloneqq   \mathbb{E}_{\BB \sim \hat{p}(\BB | \IB)}\left[\left(\frac{\BB_i}{\lVert \BB_i \rVert}\right)^{\top} \left( \frac{\BB_j}{\lVert \BB_j \rVert} \right) \right],
\end{equation}
where $C_{i,j}(\IB) \in [-1, 1]$ measures the directional agreement between magnetic field vectors at pixel $i$ and $j$. $C_{i,j}(\IB) \to 1$ means that $i$ and $j$ are correlated, $C_{i,j}(\IB) \to 0$ means they are uncorrelated, and $C_{i,j}(\IB) \to -1$ means they are anti-correlated. 
We approximate Equation~\ref{eqn:similarity} with multiple samples, akin to Monte-Carlo integration. 

We show correlation maps of representative samples in \autoref{fig:appx_correlation}. These examples show that \modelname~ captures structured nonlocal relationships in the sampled distribution, including intra-plage coherence and correlations that are largely consistent with magnetic connectivity.

This correlation structure may be useful even when the full magnetogram prediction is not used directly. For instance, given the polarity of a small set of pixels in an image from surface flux transport models \citep{surface_flux_base_1989_Wang, AFT_Upton_2014, OFT_Caplan_2025}, the learned correlations could help propagate polarity information to remaining pixels.

This ability to generate a distribution also enables estimation of uncertainty quantification (e.g., the per-pixel standard deviation), which is important for applications. Similar to the correlation map analysis, the generative nature of diffusion models allows us to draw a large ensemble of realizations and probe the per-pixel uncertainty. Before uncertainty computation, we downsample all realizations to a coarser grid. This reduces sensitivity to co-registration error, limited optical resolution, and temporal mismatch (since the UV/EUV input is near-instantaneous snapshot while the corresponding magnetogram is captured over a one-hour interval). The uncertainty is then quantified as the per-pixel standard deviation normalized by its mean magnitude, computed over the set of downsampled realizations. As an example, \autoref{fig:appx_uncertainty} shows uncertainty map for the four samples in \autoref{fig:appx_correlation}.

Our correlation map and uncertainty analysis show that models like \modelname can not only predict magnetograms, but also learn the structure of the full output distribution.

\subsection{Cross Solar Cycle Generalization}

\label{sec:cross_sc}

We next ask whether \modelname can generalize across the polarity reversal between solar cycles. This is a challenging test because, according to Hale's law, the leading polarity
in the northern and southern heliographic hemispheres reverses between solar cycles. It is therefore important to test whether \modelname, when trained only on data from one solar cycle, can generalize to other solar cycles.

To test this, we train a separate model using {\bf only observations from solar cycle $24$} and evaluate it {\bf only on solar cycle $25$ data}. During training, we use a simple polarity-flip augmentation to expose the model to both solar cycle labels while using only solar cycle 24 observations. With probability $0.5$, the dataset flips the sign of all three target vector magnetogram components and changes the solar cycle indicator to the solar cycle $25$ value, while leaving the UV/EUV inputs and latitude channel unchanged. This augmentation provides examples that mimic solar cycle $25$ and helps address the cross cycle generalization difficulty. Apart from this augmentation, training follows the standard \modelname procedure. 

At test time, solar cycle $25$ samples spanning 2020 to 2024 are passed into the model with their true solar cycle indicator. Using the polarity score defined in \autoref{eqn:polarity_ratio}, the model recovers the correct polarity for $89.7\%$ among $16,201$ samples with threshold $\tau < 1$. With a stricter threshold $\tau < 0.5$, we get a passing rate of $83.3\%$. Beyond polarity agreement, qualitative inspection shows performance consistent with the original \modelname model. These results indicate that \modelname can generalize across solar cycles.

\subsection{Full Disk Generalization}
\label{sec:full_disk_generalization}

\begin{figure*}[t]
    \centering
    \includegraphics[width=1\linewidth]{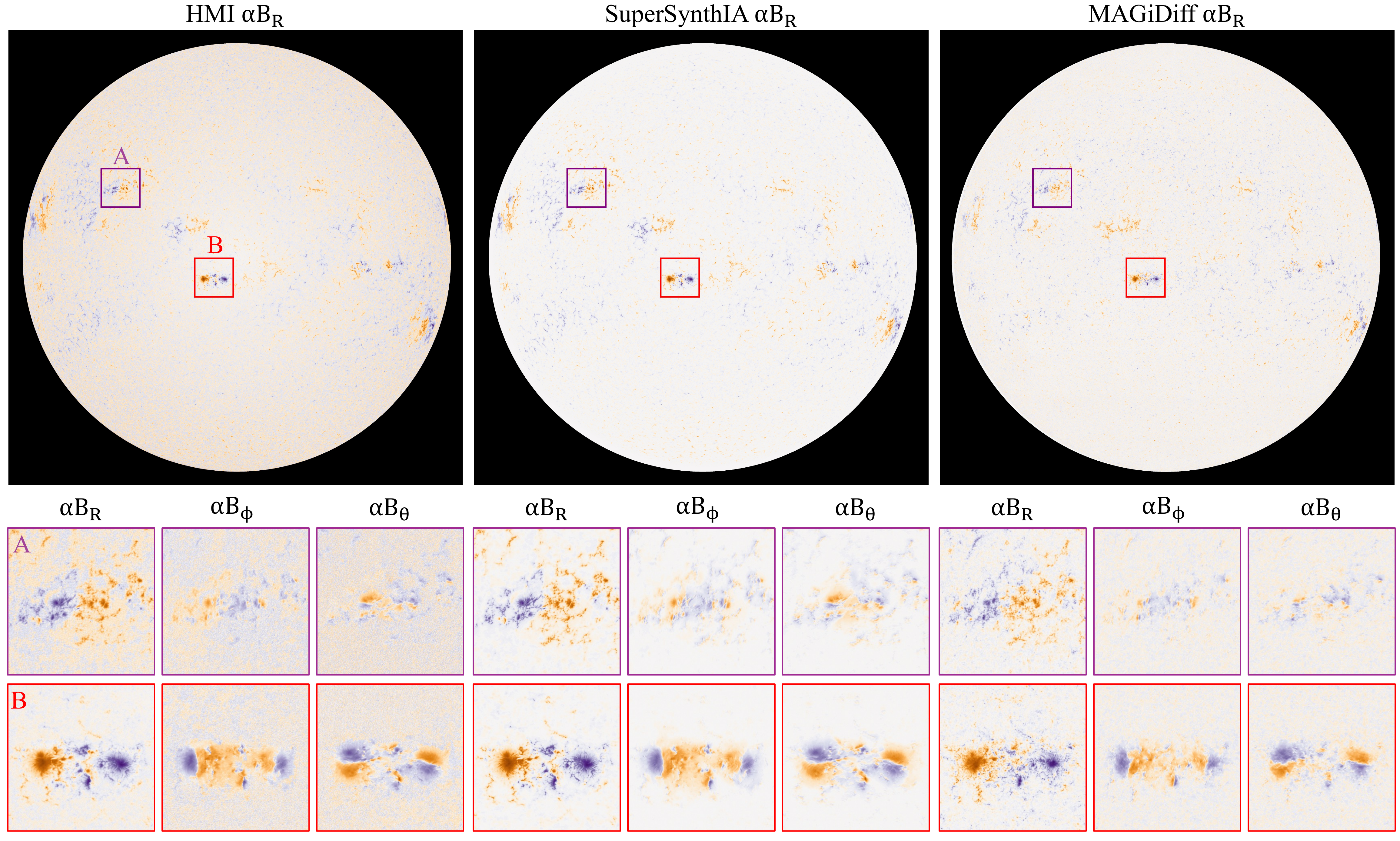}
    \caption{\textbf{Full-disk examples from 2016 February 5, 07:12 TAI.} 
    \textbf{Upper Panel:} Left to right:
    $\alpha B_R$ for \hmi, SuperSynthIA, \modelname. 
    \textbf{Bottom Panel:} Left to right: $\alpha B_R$, $\alpha B_\phi$, $\alpha B_\theta$ for region A and B. 
    The magnetic field structure inferred by  \modelname largely mimics those of HMI and SuperSynthIA despite a preferential direction in some large-scale plage. We attribute this artifact to the patch-to-full disk domain gap: \modelname is trained only on small, activity-focused patches and thus lacks global context. As a benefit of using higher-quality \hinode magnetograms as label, \modelname quiet region more closely resembles SuperSynthIA quiet region with reduced quiet-Sun noise artifacts. Colormaps: -3000 \includegraphics[width=30pt,height=6pt]{color_PuOrSqrt.png} 3000 \gauss following \autoref{fig:fig1_qualitative}. }
    \label{fig:full_disk}
\end{figure*}

\begin{figure*}[t]
    \centering
    \includegraphics[width=1\linewidth]{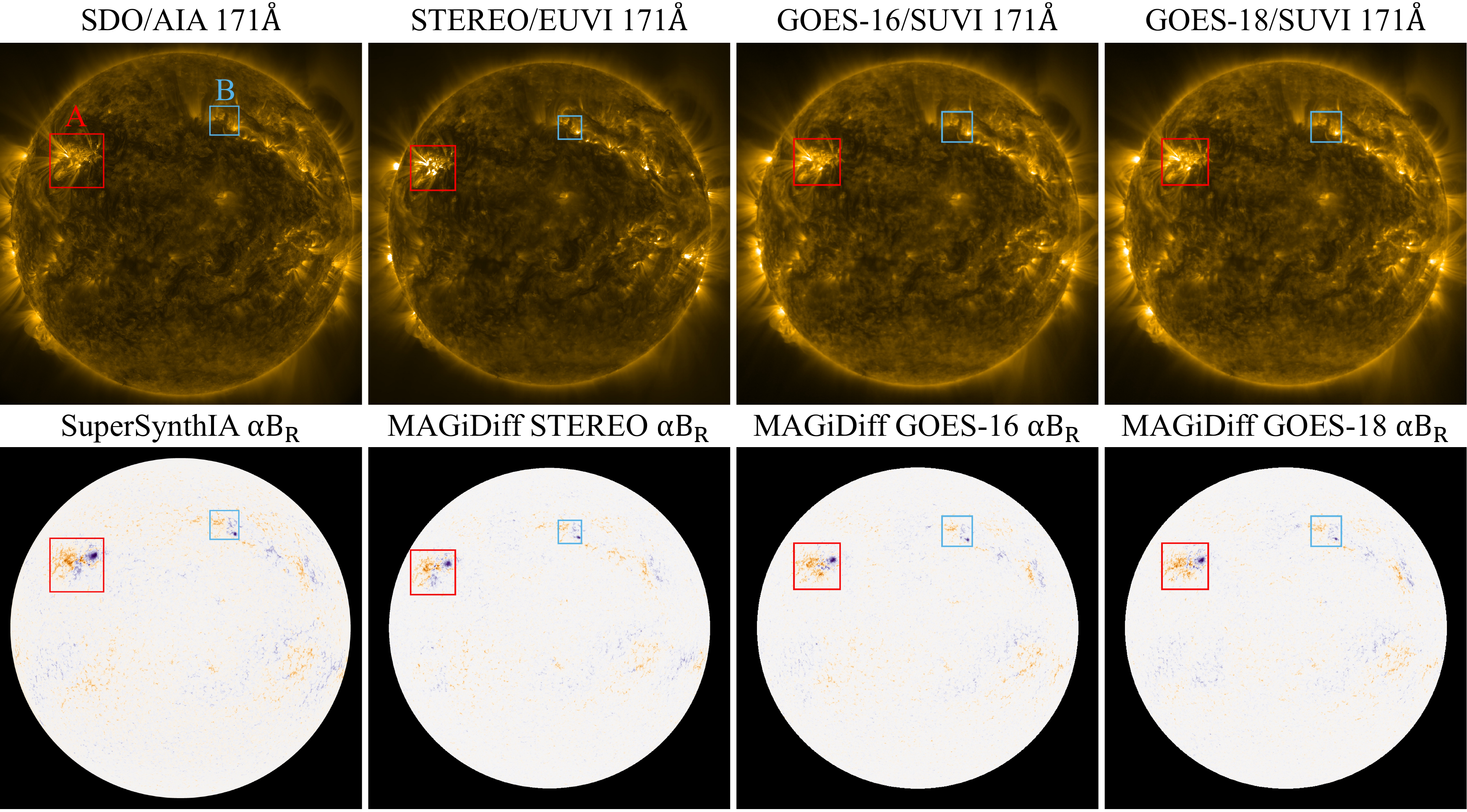}
    \caption{\textbf{Full-disk \abr predictions from \modelname across multiple EUV instruments.} 
    \textbf{Top row:} Full-disk 171\AA\ filtergrams from \aia, \stereo, \goessixteen, and \goeseighteen (left to right).
    \textbf{Bottom row:} SuperSynthIA \abr reference followed by \modelname \abr predictions using full-disk \stereo, \goessixteen, and \goeseighteen filtergrams as input (left to right).
    \modelname produces largely accurate full-disk magnetic field estimates across all three instruments, with active regions reproduced well. However, some predictions exhibit a preference for uniform polarity inconsistent with the SuperSynthIA reference, which we attribute to a domain gap arising from training exclusively on cropped regions.
   Colormaps: -3000 \includegraphics[width=30pt,height=6pt]{color_PuOrSqrt.png} 3000 \gauss following \autoref{fig:fig1_qualitative}. Example Date: 2024 February 22, 04:12 TAI. 
   }
    \label{fig:other_Instr_full_disk}
\end{figure*}

We now evaluate whether \modelname~ can be applied to full-disk UV/EUV observations. This setting is well beyond the model's training distribution: \modelname is trained on co-registered $128\arcsec \times 128\arcsec$ cutouts, whereas full-disk inference requires the model to handle much larger spatial context, extended plage regions, and the off-disk area. Therefore, some aspects of full-disk prediction of magnetograms are {\it expected} to not work — e.g., large-scale plage patches spanning larger regions. Nevertheless, it is unclear whether the method would generalize even to active regions. Indeed, in early development, we found that \modelname~{\it without a binary disk mask} produced spurious magnetogram hallucinations near disk center when applied directly to full-disk inputs.

In addition to the binary disk mask, we find that two modifications are critical for producing spatially coherent full-disk predictions. 
First, motivated by resolution-dependent noise scheduling~\citep{hoogeboom_simple_2023, esser_scaling_2024}, we adjust the noise schedule used to train \modelname. As image resolution increases, a large-scale structure spans more pixels and thus the same per-pixel noise level has a weaker effect on the structure since the independent pixel noise averages out over the large number of pixels of the structure. Consequently, a noise schedule tuned on $128\arcsec \times 128\arcsec$ crops is too weak on full-disk images. We therefore retrain \modelname from scratch with the log signal-to-noise ratio lowered by $3.65$ nats at every diffusion timestep. This shift corresponds to a roughly $6$ times larger canvas, making the denoising difficulty more comparable between training and inference time.
Second, we adopt a tiled inference procedure motivated by MultiDiffusion~\citep{bartal2023multidiffusionfusingdiffusionpaths}. As in standard diffusion inference, the process begins with a full-disk canvas of random noise and progressively denoises it into a clean magnetogram. The difference is that, at each denoising step, the canvas is divided into overlapping  $256\arcsec \times 256\arcsec$ windows, and \modelname predicts a denoising update for each window using the corresponding UV/EUV observation crop. These updates are averaged in overlapping regions and combined to form a full-disk estimate used in the next denoising step. This procedure allows \modelname to operate at a spatial scale close to what it is trained on while producing a spatially coherent full-disk prediction.

As a demonstration of this approach, we show \modelname prediction on a full-disk map from 2016 February 5, 07:12 TAI, which lands in the test set time range. We compare against two references. The first is the standard \hmi vector magnetogram product. The second is SuperSynthIA~\citep{wang_supersynthia_2024}, which uses \hmi Stokes profiles to produce disambiguated magnetograms that resemble \hinode. SuperSynthIA and \hmi agree in many places, but have some differences in quiet regions and plage characteristics. 

We show results in Figure~\ref{fig:full_disk}. \modelname~ recovers active-region magnetic structure that broadly resembles both \hmi and SuperSynthIA. 
The hexbin plots in Figure~\ref{fig:full_disk_hexbin} provide a quantitative comparison with the SuperSynthIA reference: the dominant pixel densities concentrate near the $1{:}1$ relation across all field components. Although the slopes of fitted lines are below unity and might suggest systematic field strength underestimation, closer inspection shows that this disagreement is not uniform across all field strengths. Strong field pixels generally remain close to the $1{:}1$ relation, whereas the dominating weak and intermediate field pixels are more significantly underestimated. These pixels mainly correspond to plage and weak fields, which are more difficult to recover as their magnetic signals are weaker, more spatially diffused, and less constrained by the UV/EUV observations. 

The remaining failure occurs primarily in large plage regions, where \modelname may assign the wrong polarity to part of a bipolar structure and may produce overly coherent unipolar structure. Although the location and morphology of the plage are preserved, its polarity can be incorrect.
We hypothesize that this stems from the training setup: at the $128\arcsec \times 128\arcsec$ scale, plage regions often appear as locally unipolar patches, so the model has limited context for resolving large-scale plage polarity.

Overall, these results suggest that \modelname can generalize beyond \hinode-sized cutouts, but also highlight the need for additional global context when resolving large-scale structure. In practical applications, this limitation could be mitigated in several ways. One approach is to take polarity from other signals for plage (e.g., in a far-side estimation task, using data from a surface flux transport model). Another approach is to use a hierarchical model that produces global polarity information (e.g., a variant of \modelname trained to map from \aia to \hmi).

\subsection{Migration to \stereo and \goes Data}

\begin{figure*}[tp]
    \centering
    \includegraphics[width=1\linewidth]{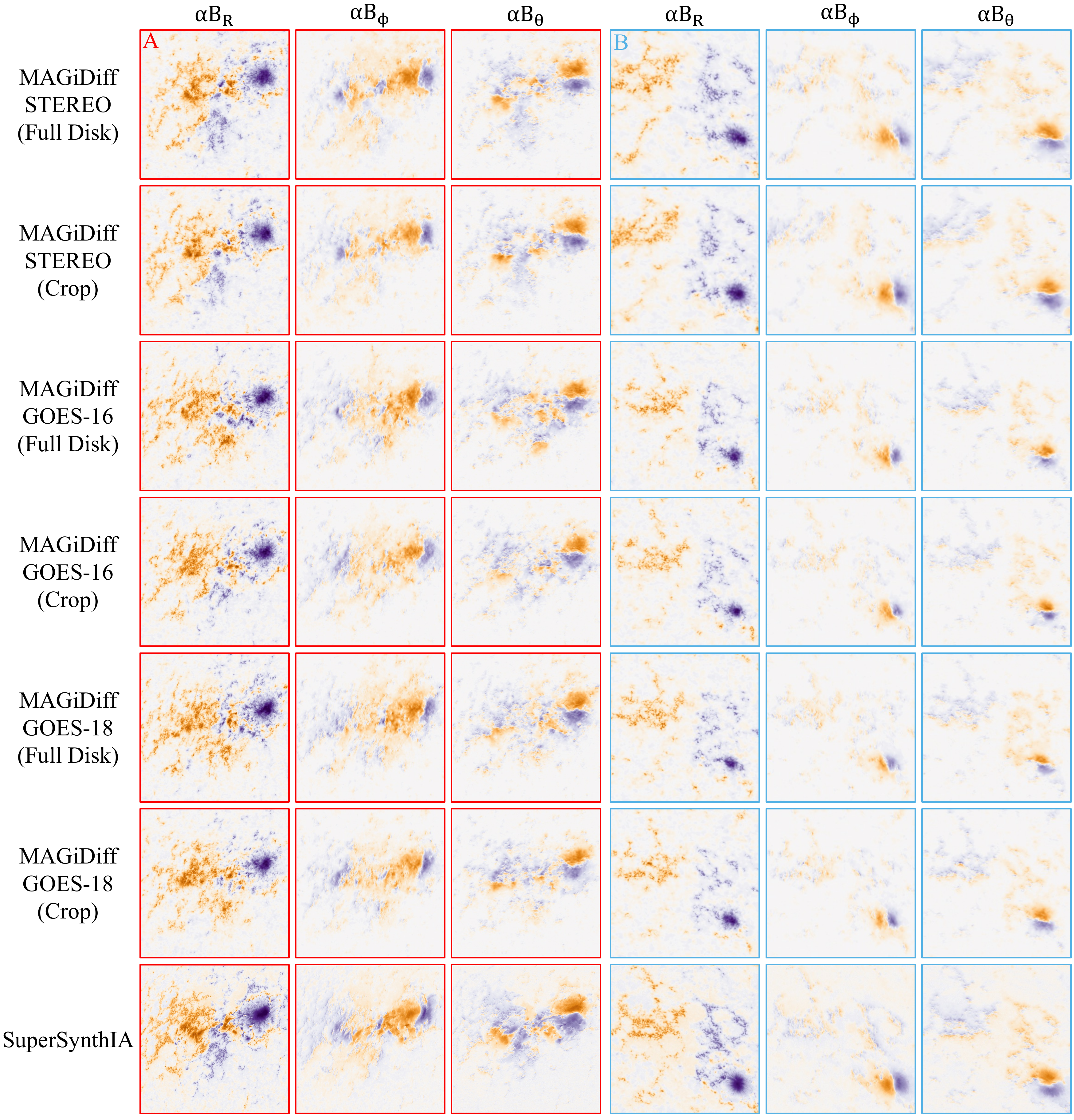}
    \caption{\textbf{Zoomed-in view of cutouts from \autoref{fig:other_Instr_full_disk}.} 
    Left to right: $\alpha B_R$, $\alpha B_\phi$, $\alpha B_\theta$ for cutout region \textbf{A} and \textbf{B}. 
    Top to bottom: \modelname predictions from $3$ channel full-disk \stereo EUV filtergrams, cropped post-inference to regions \textbf{A} and \textbf{B} (row 1), and inferred directly on the EUV cutouts of the same regions (row 2); same for $5$ channel \goessixteen EUV filtergrams (rows 3 -- 4) and $5$ channel \goeseighteen EUV filtergrams (rows 5 -- 6); SuperSynthIA reference (row 7).
    Across all three instruments, \modelname recovers the dominant magnetic field structures under both inference modes, and the predictions closely resemble the SuperSynthIA reference. 
   Colormaps: -3000 \includegraphics[width=30pt,height=6pt]{color_PuOrSqrt.png} 3000 \gauss following \autoref{fig:fig1_qualitative}. Example Date: 2024 February 22, 04:12 TAI. 
   }
    \label{fig:otherInstr_cutout}
\end{figure*}

A primary motivation for \modelname is expanding the scope of data that can be obtained by using instruments that lack a spectropolarimeter. We thus test whether \modelname can be adapted to two such instruments: the Extreme Ultraviolet Imager onboard \textit{STEREO} (\stereo) and the Solar UltraViolet Imager onboard \textit{GOES-R} (\goes). Both deliver multi-wavelength EUV imagery, but neither has the full \aia channel set, and both differ from \aia in photometric calibration, spatial resolution, and point-spread function. In addition, \stereo observes the Sun from a substantially different vantage point.

This setting is more challenging than the \aia-to-\hinode task for two main reasons. 
First, the available EUV passbands primarily originate in the chromosphere, transition region, and corona rather than the photosphere, thus provide only indirect constraints on the photospheric magnetic field. 
Second, neither \textit{STEREO} nor \textit{GOES} provides vector magnetogram measurements, and \hinode does not provide sufficiently dense co-observations to construct paired training sets.  
Because true \stereo far-side predictions cannot be directly validated without simultaneous far-side vector magnetograms, we evaluate the model in a proxy setting in which magnetic supervision is available: we use periods when each EUV instrument has substantial field-of-view overlap with \hmi, generate SuperSynthIA vector magnetograms from the corresponding \hmi Stokes observations, and reproject these magnetograms onto the coordinate grid of the respective instrument.
These reprojected SuperSynthIA magnetograms are used as the fine-tuning target and the validation reference.
This proxy experiment provides a way to validate whether \modelname can generalize to other EUV instruments and recover vector magnetograms from their EUV observations.

We adapt \modelname to each instrument using a two-stage fine-tuning procedure. First, we fine-tune the VAE on SuperSynthIA magnetograms so that the latent space better matches the target magnetogram distribution. Second, we fine-tune two separate denoising U-Nets for \stereo and \goes on their corresponding instrument-specific dataset while keeping the VAE frozen.
The base model that is fine-tuned upon is trained with the resolution-adjusted noise schedule described in Section~\ref{sec:full_disk_generalization}. The resulting full-disk prediction generation also follows the tiled inference procedure.
The detailed fine-tuning procedure is provided in Section~\ref{sec:appx_cross_instr_fine_tuning}. 

We show qualitative results in \autoref{fig:other_Instr_full_disk}, which presents full-disk \abr predictions from the fine-tuned \stereo and \goes models. We evaluate the models on data from 2024 February 22, 04:12 TAI, which falls within the test-set time range and is selected because the \stereo field of view has substantial overlap with both \hmi and \goes. Across all three instruments, \modelname recovers the dominant active-region magnetic structure and produces full-disk predictions that broadly resemble the SuperSynthIA reference.
Some disagreement remains in extended plage regions, where a bipolar structure may be predicted as unipolar, consistent with the limitation discussed in Section~\ref{sec:full_disk_generalization}.

We additionally assess the distributional agreement between these full-disk predictions and the SuperSynthIA reference in \autoref{fig:other_Instr_full_disk_hist}. Because the predictions are inferred solely from EUV observations acquired by different instruments, whereas the SuperSynthIA reference is derived from \hmi observations and reprojected onto each instrument's respective coordinate grid, exact pixel-accurate correspondence is impossible. We therefore examine the pixel-value distributions to assess whether the models recover the overall distribution of magnetic field values. Across all instruments and field components, the predicted distributions broadly follow the corresponding SuperSynthIA distribution, with slight underestimation at strong field strengths, particularly for \abp and \abt.

To isolate the effect of full-disk domain gap, we compare two inference settings in \autoref{fig:otherInstr_cutout}: applying \modelname to the full-disk EUV input and then cropping the predicted magnetogram afterward, versus applying \modelname directly to the corresponding EUV cutout. With tiled full-disk inference, the two settings produce similar results in both regions, and both recover the dominant magnetic structures well.
The zoom-ins also explain the  underestimation of strong-field \abp and \abt pixels for the two \goes predictions in \autoref{fig:other_Instr_full_disk_hist}. In both inference settings, \modelname predicts a smaller sunspot than SuperSynthIA, resulting in substantial field strength underestimation for pixels near the sunspot boundary. This underestimation of the size of the active region is not too surprising, since the \goes instruments only provide EUV information and have limited direct information about the photosphere. Nonetheless, the results demonstrate that tiled inference can extend the method across the full disk. More broadly, while the {\it size} of the active region in the \goes data is smaller, its general shape and configuration are consistent with SuperSynthIA. Thus, the results suggest that \modelname can be applied to a variety of UV/EUV instruments to produce estimates of the vector magnetic field.

\section{Discussion} \label{sec:discussion}
We present \modelname, a new approach to estimate photospheric vector magnetic field directly from UV/EUV intensity maps using a deep learning model. \modelname is built upon the latent diffusion model and is trained on co-registered \aia and \hinode observations with $14$ years of data. Inspired by prior works \citep{kim_solar_2019, jeong_solar_2020, sun_dynamic_2022}, we extend the focus beyond line-of-sight magnetograms to the full heliographic vector magnetic field components \abr, \abp, \abt. Qualitative and quantitative evaluations show that \modelname produces visually-plausible magnetograms. In cutouts, \modelname predictions show strong agreement with \hinode, capturing both the large-scale structure and fine details. 

We propose methods to probe the model beyond standard prediction to examine the richer physical information embedded in the model's predictive distribution. We further demonstrate that \modelname can be adapted to other EUV instruments and shows promising generalization across solar cycles. Full-disk experiments show that \modelname can recover the main active-region magnetic structure beyond \hinode-sized cutouts, while also revealing the need for additional global context to resolve domain gaps.

Many of these results are credited to the breadth of data that is available for learning methods. \hinode has provided nearly 19 years of high-quality data, and \aia has provided high-cadence, well-calibrated, full-disk filtergrams since 2010. This immense archive of data is absolutely critical for creating a high-quality dataset for training.

Meanwhile, deep learning has unlocked huge potential in solar physics. Traditionally, vector magnetograms can only be inferred using inversion methods like \cite{Lites_MERLIN_2007, borrero2011vfisv}, which require high quality Stokes profiles. These full Stokes profiles are hard to acquire as a spectropolarimeter is required and typically can only be measured with limited cadence and coverage. By contrast, EUV intensity imaging is easier and therefore more common and spans over multiple solar cycles. 
Several studies have demonstrated that models can reasonably infer magnetograms from UV/EUV intensity images, despite the only statistical relationship between UV/EUV intensity and magnetic field maps. Among these efforts, \cite{kim_solar_2019} was the first to propose using a conditional GAN to estimate LOS magnetogram from EUV intensity images. Subsequent works - \citep{jeong_solar_2020, deng_improving_2021, sun_dynamic_2022, jiang2023generating, gao_generating_2023, li_transfer_2024, jarolim_deep_2025, Jeong2025} - experimented with inputs from various instruments and introduced refinements to the GAN structure like incorporating temporal consistency and multi-channel conditioning. More recently, \citet{ramunno_solar_2024, ramunno_enhancing_2025, xu_improving_2025} bring diffusion into the field but their work primarily focuses on super-resolution of LOS magnetograms and conditional solar image generation. To the best of our knowledge, \modelname is the first diffusion-based model to estimate full heliographic vector magnetic field end-to-end from EUV observations. We further probe the model to characterize the physical correlation it learns.  We also test cross-instrument transfer to \stereo and \goes, where \modelname shows good generalization ability, suggesting a potential extension to other similar instruments with careful calibration.

Despite its strong performance, \modelname naturally inherits several limitations due to its data-driven methodology. First, it depends critically on the accurate co-registration between \aia UV/EUV intensity map and \hinode magnetograms. This is difficult as \aia captures data near instantaneously while \hinode magnetograms are built up over tens of minutes. Such a temporal disparity means that UV/EUV observation can change dramatically while magnetogram data is being produced, leading to natural mismatch between the pairs. Second, \modelname is trained on small, activity-focused cutouts. When applied at full-disk scale, this crop-based training can lead to artifacts like overly coherent unipolar large-scale plage. These failure modes suggest that full-disk prediction requires additional global context beyond what is available in the training cutouts.
Third, although the model is provided with a polarity prior that should help resolve the polarity ambiguity, polarity errors can still occur. In many of these failure cases, different \modelname realizations sample both the correct polarity solution and its sign-flipped counterpart. In other cases, however, the model consistently selects the opposite polarity. These cases indicate that resolving polarity remains imperfect in the current model and could be improved in future work. 

Nonetheless, \modelname shows the opportunity to generate estimates of the vector magnetic field from other instruments. In combination with other advances in solar physics, we hope to extend the availability of high-quality data.

\begin{acknowledgments}
This work was primarily supported by NASA/MIRO  80NSSC24M0174. This work was supported in part through the NYU IT High Performance Computing resources, services, and staff expertise. The authors thank Dr.~KD Leka for a number of  helpful comments that greatly improved the manuscript. The authors thank the anonymous reviewer for insightful feedback that improved the paper.
\end{acknowledgments}

\bibliography{citation}{}
\bibliographystyle{aasjournal}

\appendix

\renewcommand*{\theHequation}
    {appendix.\Alph{section}.\arabic{equation}}

We now provide additional technical details about the diffusion models underlying \modelname. We first outline the core diffusion formulation and key techniques adopted in this work. Then, we describe the implementation and architecture details of \modelname. We conclude by presenting supplementary experiments and analyses that support our claim in the paper.

\section{Diffusion Model Details}
\label{sec:appx_diffusion_model_details}
Diffusion models lay the foundation for this work, and in this section we provide a concise overview of the underlying framework. We begin with an introduction of the classical formulation of the forward and reverse processes in the original denoising diffusion probabilistic model (DDPM, \cite{ho_denoising_2020}), which defines the basic mechanism of gradually adding and removing noise. Building on that, we then introduce refinements that are central to our implementation. 

\subsection{Forward and Reverse Process}
In the classical DDPM, the generative process is defined by a pair of Markov chains: a forward noising process that gradually perturbs clean data into noise, and a reverse denoising process that gradually removes noise to recover clean samples. 

\subsubsection{Forward Process}
The forward diffusion process systematically corrupts clean data with Gaussian noise and is only used during training. Starting from an unperturbed image $\xB_0 \coloneqq \xB$, Gaussian noise is gradually injected over a fixed number of steps $t \in \{1, 2, \cdots, T \}$. At each step, the transition is controlled by a predefined variance schedule $\{ \beta_t \}^{T}_{t=1}$, where $\beta_{t} \in (0, 1)$ determines how much noise is added. Formally, the forward process is a Markov chain in which the next noisy state $\xB_t$ is drawn from a Gaussian distribution with mean centered on $\sqrt{1 – \beta_t} \xB_{t-1}$ with variance $\beta_t$:

\begin{equation}
\label{eq:forward_step}
q(\xB_t \mid \xB_{t–1}) = \mathcal{N}(\xB_t; \sqrt{1 – \beta_t} \xB_{t–1}, \beta_t \IB)
\end{equation}
For simplicity, this can be rewritten in closed form as:
\begin{equation}
    \xB_t = \sqrt{\overline{\alpha}_t} \xB_0 + \sqrt{1 - \overline{\alpha}_t} \epsilonB
    \label{eq:forward_closed}
\end{equation}
where it resembles the Markov process in a single step. The noisy image $\xB_t$ can thus be written directly as a linear combination of the clean image $\xB_0$ and Gaussian noise  $\epsilonB \sim \mathcal{N} (0, I)$, with weights determined by $\overline{\alpha}_t \coloneqq \Pi_{s=1}^{t} 1-\beta_s$.

Intuitively, as $t$ increases, $\xB_t$ contains less information from $\xB_0$ while more dominant by noise. In the limit $t \rightarrow T$, $\xB_t$ approaches pure Gaussian noise with no information about the original clean image.

\subsubsection{Reverse Process}
While the forward diffusion process gradually destroys information in the image, the reverse process defines how to reconstruct clean images from noise. Because the forward process is Gaussian, the reverse process also has an exact Gaussian form: 
\begin{equation}
    \label{eq:reverse_exact_gaussian}
    q(\xB_{t-1} \mid \xB_t, \xB_0) = \mathcal{N}(\xB_{t-1} ; \tilde{\muB}_t(\xB_t, \xB_0), \tilde{\beta}_t \IB)
\end{equation}
with the mean $\tilde{\muB}_t(\xB_t, \xB_0)$ and variance $ \tilde{\beta}_t$ defined as:
\begin{equation}
    \tilde{\muB}_t(\xB_t, \xB_0) \coloneqq \frac{\sqrt{\overline{\alpha}_{t-1} }\beta_t}{1-\overline{\alpha}_t} \xB_0 + \frac{\sqrt{\alpha_t} (1- \overline{\alpha}_{t-1})}{1-\overline{\alpha}_t} \xB_t  \qquad
     \tilde{\beta}_t \coloneqq \frac{1 - \overline{\alpha}_{t-1}}{1-\overline{\alpha}_t} \beta_t
\end{equation}
However, in practice this is intractable, since the clean image $\xB_0$ is unavailable at inference time and thus the mean $\tilde{\muB}_t(\xB_t, \xB_0)$ cannot be computed.

Instead, DDPM approximates this reverse chain by making the network predict the noise $\epsilonB$ that was added in the forward process, and then recovering an estimate of $\xB_0$ using \autoref{eq:forward_closed}. Let $f_{\theta}$ denote the denoising diffusion network with learnable parameters $\theta$. Given a noisy image $\xB_t$ at timestep $t$ and the timestep $t$, the network outputs the predicted noise, which leads to the estimate of $\xB_0$:

\begin{equation}
    \label{eq:x0_approx}
    \hat{\xB}_0 = \frac{1}{\sqrt{\overline{\alpha}_t}} (\xB_t - \sqrt{1 - \overline{\alpha}_t} f_{\theta} (\xB_t, t))
\end{equation}
This estimated clean image $\hat{\xB}_0$ can then be substituted back into \autoref{eq:reverse_exact_gaussian}, yielding an update rule that maps $\xB_t$ to a cleaner image $\xB_{t-1}$ defined as:
\begin{equation}
    \label{eq:DDPM_update_rule}
    \xB_{t-1} = \frac{1}{\sqrt{\alpha_t}} (\xB_t - \frac{\beta_t}{\sqrt{1 - \overline{\alpha}_t}}  f_{\theta} (\xB_t, t)) + \sigma_t \etaB
\end{equation}
where $\etaB \sim \mathcal{N}(0, \IB)$ is standard Gaussian noise and sampling variance $\sigma_t$ is typically chosen as $\sigma_t^{2} =  \tilde{\beta}_t$. The stochastic term $\sigma_t \etaB$ injects a small amount of noise to help maintain diversity in the reverse process.

Training is therefore performed by minimizing the mean-squared error between the true noise and the network predicted noise at randomly sampled timesteps $t \sim \text{Uniform}\big(\{1,2,\ldots,T\}\big)$.
\begin{equation}
    \mathcal{L} = \mathbb{E }_{\xB_0, \epsilonB, t} \| \epsilonB - f_\theta (\xB_t, t) \|^{2}
\end{equation}
This objective teaches the network to provide accurate denoising predictions all noise levels. During inference time, starting from pure Gaussian noise $\xB_T \sim \mathcal{N}(0,1)$, the reverse process is applied iteratively to predict noise and update the sample, gradually transforming noise into a clean image. 

\subsection{$v$-prediction parameterization}
In addition to directly predicting the noise $\epsilonB$ or the clean image $\xB_0$, \cite{v_obj_Salimans} introduced an alternative parameterization known as $v$-prediction, which we adopt in \modelname. In this formulation, the model predicts a specific linear combination of $\epsilonB$ and $\xB_0$, which we define as the velocity term $\vB$:
\begin{equation}
    \label{eq:v_prediction_definition}
    \vB = \sqrt{\overline{\alpha}_t}\epsilonB - \sqrt{1-\overline{\alpha}_t} \xB_0
\end{equation}
This reparameterization is invertible as given $(\xB_t, \vB)$, according to \autoref{eq:forward_closed}, one can recover both $\xB_0$ and $\epsilonB$. Thus, predicting $v$ is mathematically equivalent to predicting $\epsilonB$ or $\xB_0$, but with a weighting that yields a better training signal across timesteps as gradients are less dominated by the high-noise or low-noise extremes, improving optimization and sampling quality. The training loss therefore is calculated as:
\begin{equation}
    \mathcal{L} = \mathbb{E }_{\xB_0, \epsilonB, t} \| \vB - f_\theta (\xB_t, t) \|^{2}
\end{equation}

\subsection{DDIM sampling}
Whereas DDPM reverse process defines a stochastic Markov chain with Gaussian noise injected at every denoising step, \cite{song_denoising_2022} introduced Denoising Diffusion Implicit Models (DDIM): a deterministic, non-Markovian sampling procedure that uses the same trained denoising network but eliminates per-step noise injection. DDIM also preserves the same distribution of $\xB_t$ at each timestep as DDPM, while enabling step-skipping and faster sampling. 

Since every noisy state can be expressed as a linear combination of the clean image $\xB_0$ and forward process added noise $\epsilonB$, the key idea of DDIM is to use the network's predicted noise $f_\theta (\xB_t, t)$ to estimate the clean image $\hat{\xB}_0$ (via \autoref{eq:x0_approx}) and then recompose a less noisy state $\xB_s$ at any earlier timestep $s < t$:
\begin{equation}
    \label{eq:DDIM_update_rule}
    \xB_s \;=\; \sqrt{\bar\alpha_s}\,\hat{\xB}_0 \;+\; \sqrt{1-\bar\alpha_s}\,f_\theta(\xB_t,t)
\end{equation}
This update rule differs from the DDPM (\autoref{eq:DDPM_update_rule}) in that no additional Gaussian noise is injected. As a result, the reverse trajectory from noise to data becomes deterministic and randomness comes from the initial noise only.

In practice, this deterministic formulation allows subsampling of timesteps. Instead of performing the full chain of $T \sim 1000$ steps in DDPM, one may select a coarser sequence of $K \ll T$ steps (e.g., $K \approx 50$) and apply the DDIM update rule, substantially reducing the cost of generation while maintaining sample quality.

\subsection{Latent Diffusion}
Applying diffusion directly in pixel space is computationally demanding, as high-dimensional images require large memory and long training times. In addition, pixel space encodes redundant high-frequency variations and sensor noise that are perceptually uninformative, forcing the model to waste capacity on irrelevant details. 

To address these challenges, Latent Diffusion Models (LDM; \cite{rombach_high-resolution_2022}) first compress images into a lower-dimensional latent space using a variational autoencoder (VAE; \cite{VAE_Kingma_2013}), and then perform the forward and reverse diffusion process in that latent space. A VAE is an encoder-decoder structure: the encoder $\mathcal{E}$ maps input images $\xB$ into a compressed latent representation $\zB = \mathcal{E}(\xB)$, and the decoder $\mathcal{D}$ reconstructs the image from its latent $\hat{\xB} = \mathcal{D}(\zB)$. The autoencoder is trained independently first using reconstruction loss and regularization, ensuring that the latent representation preserves semantic content while remaining smooth. 

Once the VAE is trained and frozen, the diffusion processes operate entirely in latent space, with the image variable $\xB$ replaced by its latent representation $\zB$. For example, the closed-form forward process (\autoref{eq:forward_closed}) now becomes:
\begin{equation}
    \label{eq:LDM_forward}
    \zB_t = \sqrt{\overline{\alpha}_t} \zB_0 + \sqrt{1 - \overline{\alpha}_t} \epsilonB
\end{equation}
and the reverse process follows the same update rule, yielding a denoised latent $\hat{\zB}_0$ that is finally decoded into image space by the decoder $\hat{\xB} = \mathcal{D}(\hat{\zB}_0)$.

This latent diffusion formulation combines the representational power of autoencoders with the generative flexibility of diffusion models. By shifting the diffusion process into a compressed latent space, it reduces computational cost, improves scalability to high-resolution data, and enhances model's knowledge on semantically meaningful information. As a result, LDMs achieve better efficiency and robustness, making it a good foundation for \modelname.

\section{Training and Inference Process Walkthrough}
\label{sec:appx_ldm_walkthrough}

In this section, we explain the workflow of \modelname, including the training of VAE (Table \ref{tab:appx_vae_training}), training of denoising network (Table \ref{tab:appx_unet_training}), and the inference process (Table \ref{tab:appx_full_inference_process}). We also show input, output, and output sizes of each step in Tables \ref{tab:appx_vae_training}, \ref{tab:appx_unet_training}, \ref{tab:appx_full_inference_process}.
\FloatBarrier

\begin{deluxetable}{lccc}
\tablewidth{0pt}
\tablecaption{Training Process of VAE}
\label{tab:appx_vae_training}
\tablehead{
\colhead{Operation} & \colhead{Input} & \colhead{Output} & \colhead{Output Shape}
}
\startdata
Input Vector Magnetogram ($\BB$) & - & - & $N \times 3 \times H \times W$ \\
VAE Encoding & $\BB$ &  $\muB$, $\logvar$ &  $N \times 6 \times h \times w$ \\
Latent Sampling & $\muB$, $\logvar$ & $\zB$ & $N \times 6 \times h \times w$ \\
VAE Decoding & $\zB$ &  $\hat{\BB}$ &  $N \times 3 \times H \times W$ \\
Loss Calculation & $\BB$, $\hat{\BB}$, $\muB$, $\logvar$  &  $\mathcal{L}$ &  - 
\enddata
\end{deluxetable}
\onecolumngrid
\FloatBarrier

\begin{deluxetable}{lccc}
\tablewidth{0pt}
\tablecaption{Training Process of Denoising Network}
\label{tab:appx_unet_training}
\tablehead{
\colhead{Operation} & \colhead{Input} & \colhead{Output} & \colhead{Output Shape}
}
\startdata
Input Vector Magnetogram ($\BB$) & - & - & $N \times 3 \times H \times W$ \\
VAE Encoding & $\BB$ &  $\zB$ &  $N \times 6 \times h \times w$ \\
Forward Noising Process & $\zB$, $\epsilonB$, $t$ &  $\tilde{\zB}_t$ &  $N \times 6 \times h \times w$ \\
Input Conditioning Information ($\IB$) & - & - & $N \times 11 \times H \times W$ \\
Spatial Rescaler Encoding & $\IB$ &  $\cB_{\mathrm{sp}}$ &  $N \times 11 \times h \times w$ \\
Input Solar cycle Indicator ($\sB$) & - & - & $N \times 1 \times h \times w$ \\
Concatenation of Conditioning Information  &  $\cB_{\mathrm{sp}}$, $\sB$  &  $\cB$ &  $N \times 12 \times h \times w$ \\
Concatenation of latent  &  $\tilde{\zB}_t$, $\cB$  &  $\operatorname{cat}(\tilde{\zB}_t, \cB )$ &  $N \times 18 \times h \times w$ \\
Denoising Network Prediction  & $\operatorname{cat}(\tilde{\zB}_t, \cB )$ &  $\hat{\vB}_t$ &  $N \times 6 \times h \times w$ \\
Loss Calculation & $\vB_t$, $\hat{\vB}_t$ &  $\mathcal{L}$ &  -
\enddata
\end{deluxetable}
\onecolumngrid
\FloatBarrier

\begin{deluxetable}{lccc}
\tablewidth{0pt}
\tablecaption{Inference Process of \modelname}
\label{tab:appx_full_inference_process}
\tablehead{
\colhead{Operation} & \colhead{Input} & \colhead{Output} & \colhead{Output Shape}
}
\startdata
Input Noise ($\tilde{\zB}_T \sim \mathcal{N}(0,1)$) & - & - & $N \times 6 \times h \times w$ \\
Input Conditioning Information ($\IB$) & - & - & $N \times 11 \times H \times W$ \\
Spatial Rescaler Encoding & $\IB$ &  $\cB_{\mathrm{sp}}$ &  $N \times 11 \times h \times w$ \\
Input Solar cycle Indicator($\sB$) & - & - & $N \times 1 \times h \times w$ \\
Concatenation of Conditioning Information  &  $\cB_{\mathrm{sp}}$, $\sB$  &  $\cB$ &  $N \times 12 \times h \times w$ \\
Repeat For Each Timestep $t \rightarrow t_{next}$: & - & - & - \\
{[1]} Concatenation of latent  &  $\tilde{\zB}_t$, $\cB$  &  $\operatorname{cat}(\tilde{\zB}_t, \cB )$ &  $N \times 18 \times h \times w$ \\
{[2]} Denoising Network Prediction  & $\operatorname{cat}(\tilde{\zB}_t, \cB )$ &  $\hat{\vB}_t$ &  $N \times 6 \times h \times w$ \\
{[3]} Diffusion Scheduler Update  & $\hat{\vB}_t$, $\tilde{\zB}_t$, $t$ &  $\tilde{\zB}_{t_{next}}$ &  $N \times 6 \times h \times w$ \\
VAE Decoding  &  $\hat{\zB}_0$ &  $\hat{\BB}$ &  $N \times 3 \times H \times W$
\enddata
\end{deluxetable}
\onecolumngrid
\FloatBarrier

\section{Additional Experiments Details}

For completeness, we provide additional qualitative and quantitative results that supplement the main evaluation of \modelname. We show four additional test samples from \modelname's test set; examine the stochastic predictive distribution; and describe the cross-instrument adaption experiments for \stereo and \goes, including data preprocessing, fine-tuning setup, and test set performance.

\subsection{Additional Qualitative Results}

\begin{figure}[h]
    \centering
    \includegraphics[width=1\linewidth]{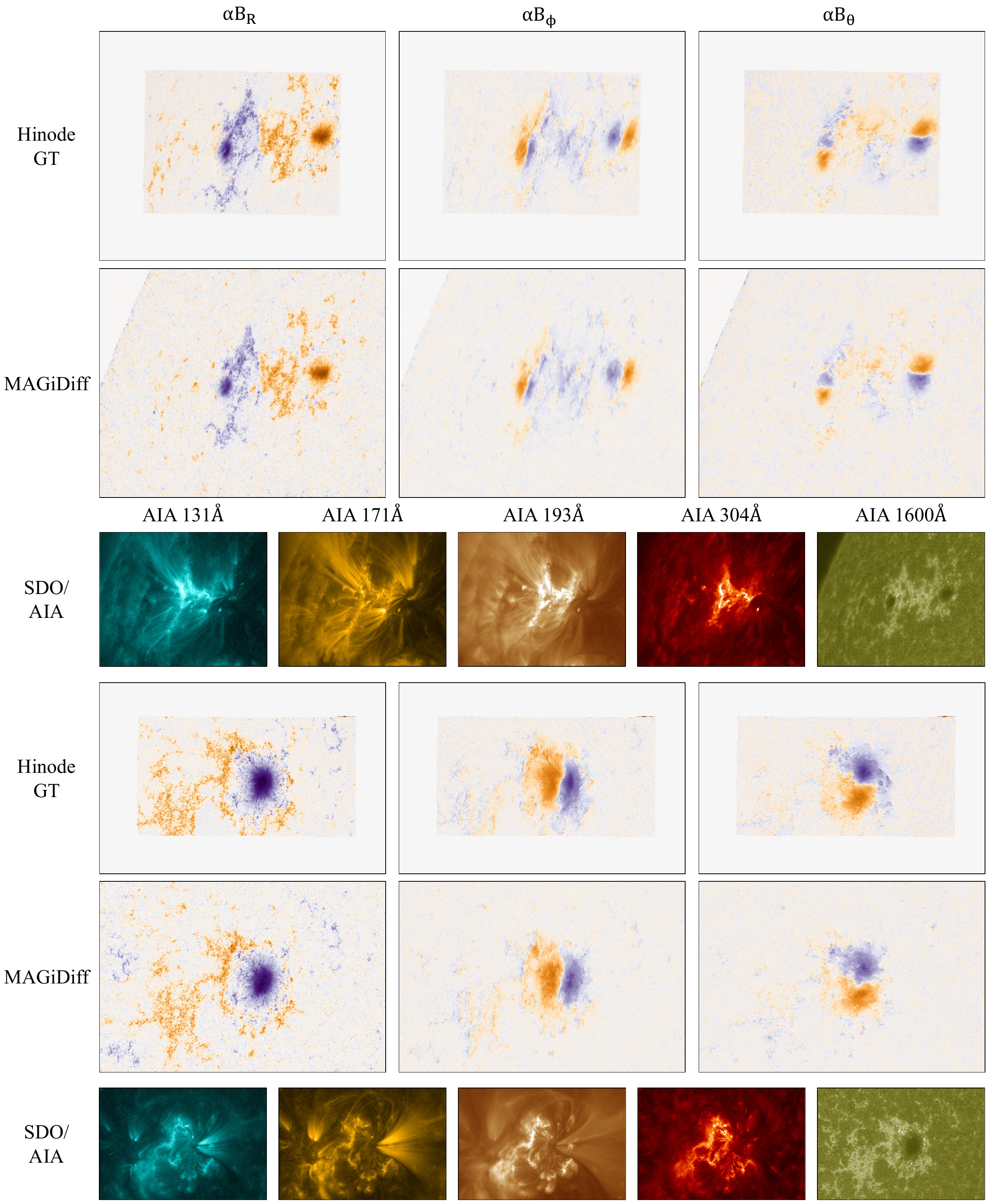}
    \caption{\textbf{Additional qualitative results for \modelname with \hinode as reference.} 
    Example date: 2016 March 18, 23:36 TAI (upper panel); 2016 May 23, 01:12 TAI (lower panel). 
    Colormaps: -3000 \includegraphics[width=30pt,height=6pt]{color_PuOrSqrt.png} 3000 \gauss following \autoref{fig:fig1_qualitative}.
     }
    \label{fig:fig1_appx_1}
\end{figure}

\begin{figure}[h]
    \centering
    \includegraphics[width=1\linewidth]{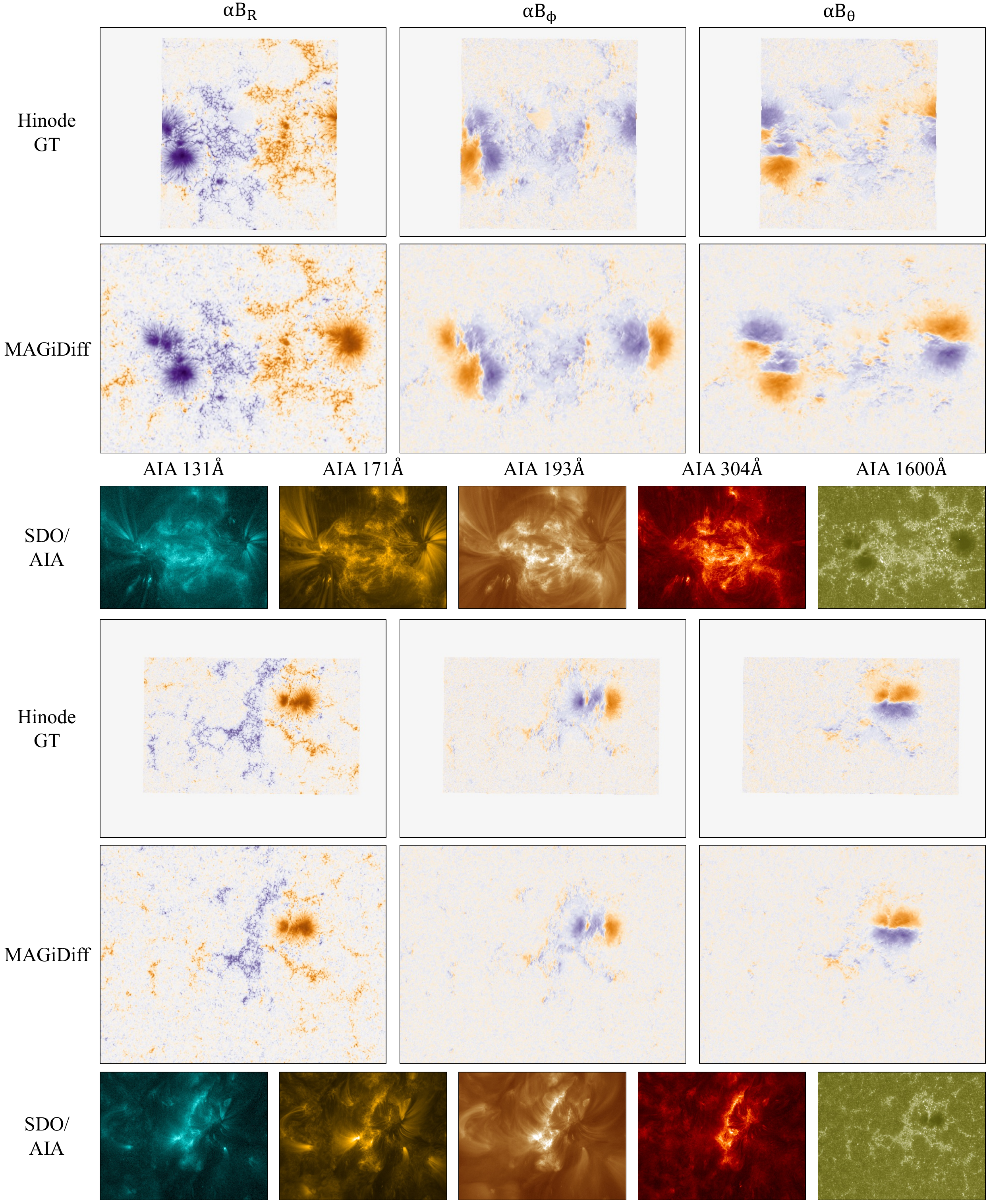}
    \caption{\textbf{Additional qualitative results for \modelname with \hinode as reference.} 
    Example date: 2016 September 5, 14:24 TAI (upper panel); 2016 November 27, 19:36 TAI (lower panel). 
    Colormaps: -3000 \includegraphics[width=30pt,height=6pt]{color_PuOrSqrt.png} 3000 \gauss following \autoref{fig:fig1_qualitative}.
     }
    \label{fig:fig1_appx_2}
\end{figure}

Here, we show some four additional panels of qualitative results for another four cutouts from the $2016$ test set in \autoref{fig:fig1_appx_1} and \autoref{fig:fig1_appx_2}. These panels show that \modelname accurately estimates \hinode like vector magnetograms at various input shapes.

\begin{figure}[h]
    \centering
    \includegraphics[width=1\linewidth]{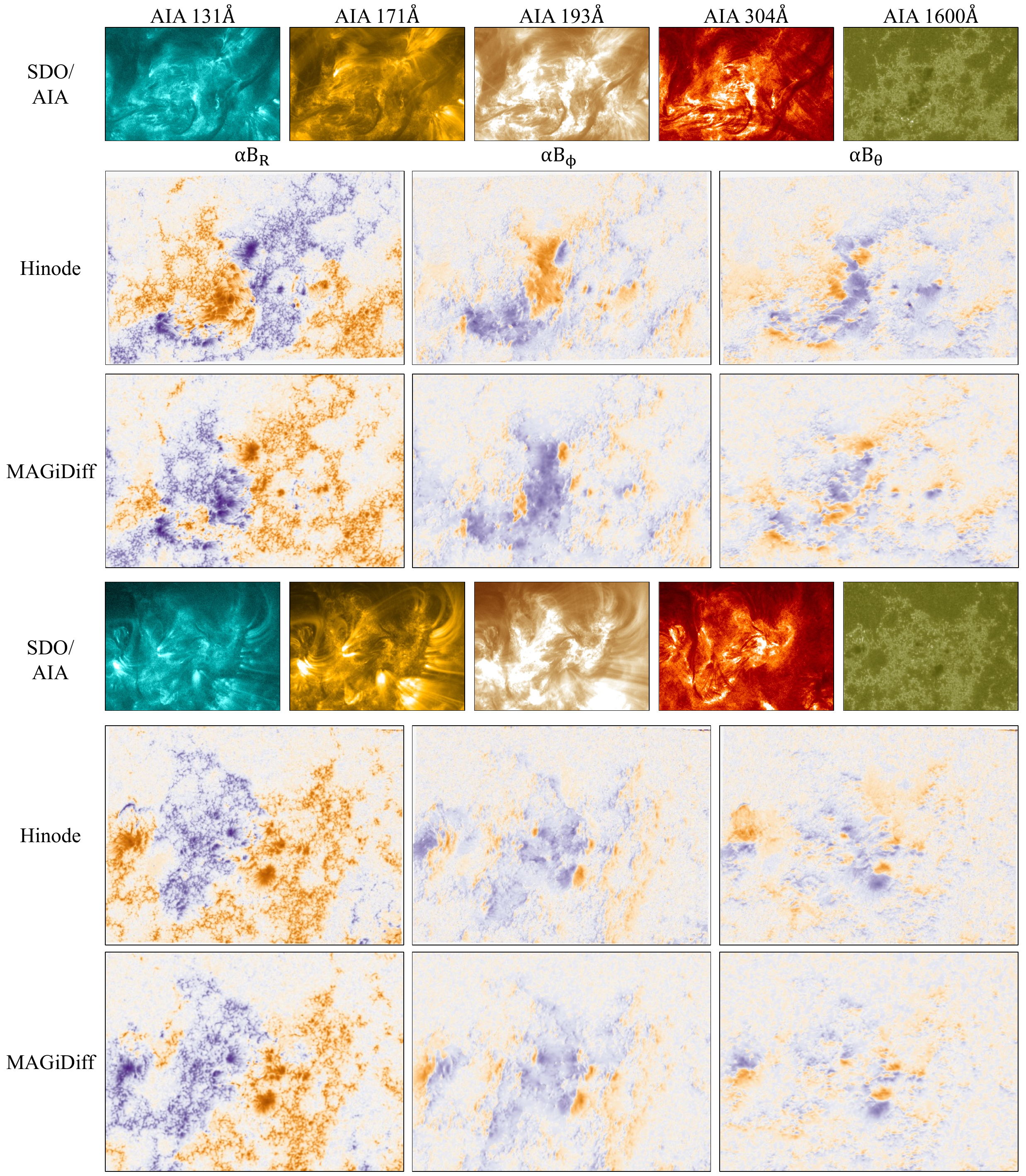}
    \caption{\textbf{Representative examples of \modelname predictions that contribute to $y=-x$ branch in \autoref{fig:hexbin}.} 
    Example date: 2024 June 5, 08:12 TAI (upper panel); 2024 July 16, 20:00 TAI (lower panel). 
    Colormaps: -3000 \includegraphics[width=30pt,height=6pt]{color_PuOrSqrt.png} 3000 \gauss following \autoref{fig:fig1_qualitative}.
     }
    \label{fig:fig_hexbin_incorrect_exp}
\end{figure}
\clearpage

\begin{figure}
    \centering
    \includegraphics[width=1\linewidth]{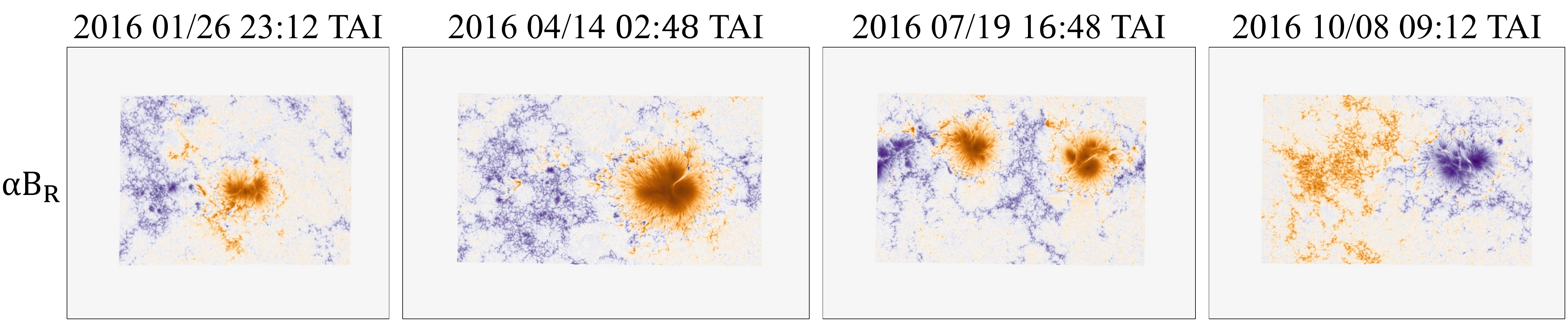}
    \caption{\textbf{Ground Truth $\alpha B_R$ for four cutouts in \autoref{fig:appx_correlation}.}
    Example dates from left to right: 2016 January 26, 23:12 TAI; 2016 April 14, 02:48 TAI; 2016 July 19, 16:48 TAI; 2016 October 8, 09:12 TAI. 
    Colormap: -3000 \includegraphics[width=30pt,height=6pt]{color_PuOrSqrt.png} 3000 \gauss following \autoref{fig:fig1_qualitative}.
   }
    \label{fig:appx_correlation_gt}
\end{figure}

\subsection{Additional Stochastic Sampling Results}

The evaluations presented in Section~\ref{sec:baseline comparison} use one realization per input in order to compare \modelname directly with the deterministic regression baseline. However, as a conditional diffusion model, \modelname encodes a distribution of plausible magnetograms and is inherently stochastic: for a fixed UV/EUV conditioning input, different initial noise can yield different realizations drawn from the learned distribution. In this section, we examine this distribution through qualitative comparisons of independent realizations and a quantitative best-of-$k$ analysis.

\subsubsection{Qualitative Comparison between Realizations}

\begin{figure*}
    \centering
    \includegraphics[width=1\linewidth]{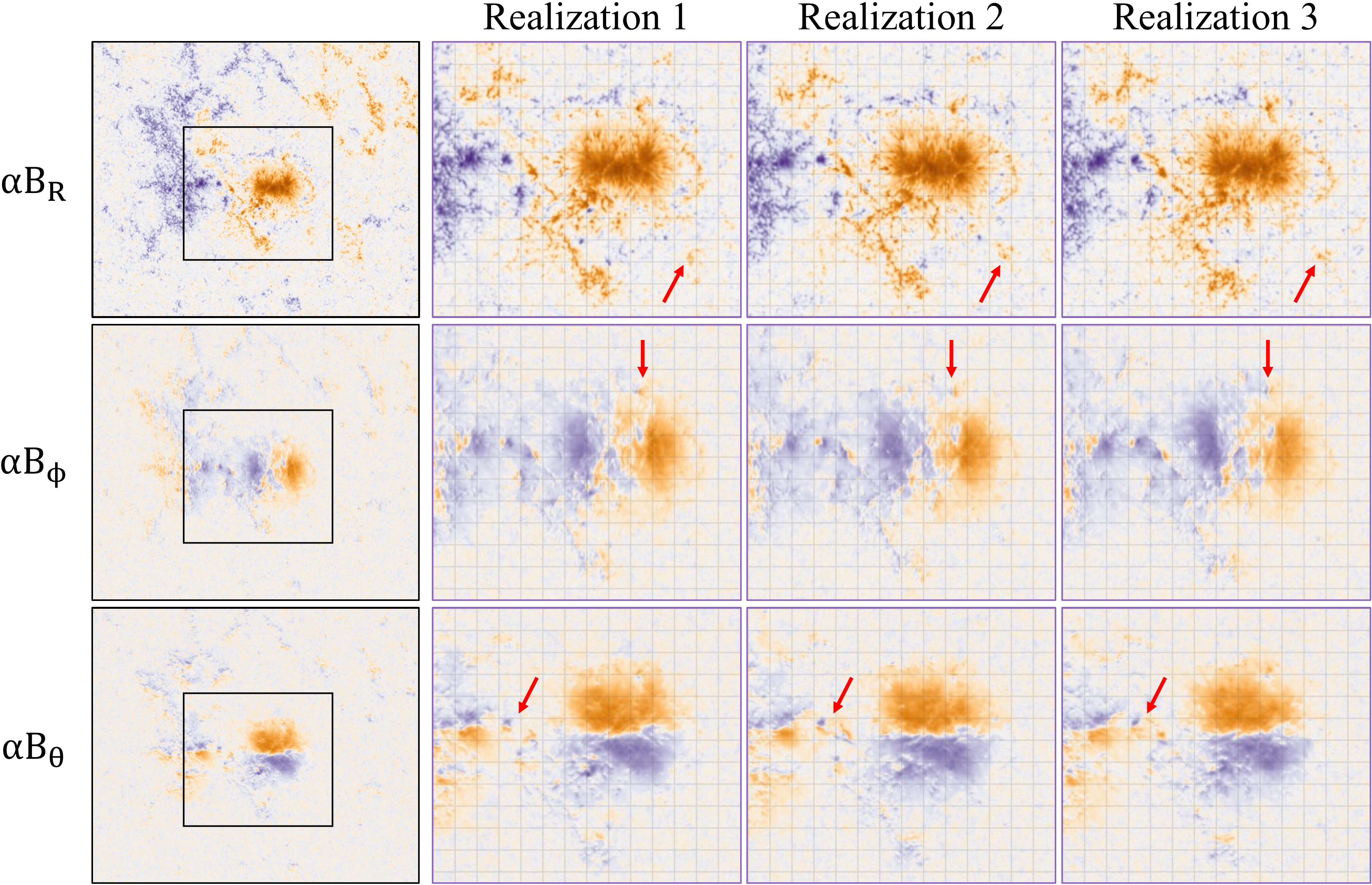}
    \caption{
    \textbf{Consistency of \modelname across multiple realizations for a single input.}
      The leftmost column shows the full-field \modelname prediction for \abr, \abp, \abt, with black boxes indicating the zoomed regions. The remaining columns show zoomed-in predictions from three independent realizations generated from the same input. The realizations show consistent magnetic structure and polarity across all three components, while retaining small-scale stochastic variations, an example of which is marked by red arrows. This example demonstrates that, for a fixed input, \modelname defines a predictive distribution from which multiple plausible samples can be drawn. These samples remain largely consistent with one another in well-constrained regions, while allowing localized variation where signal is weak.
    Colormaps: -3000 \includegraphics[width=30pt,height=6pt]{color_PuOrSqrt.png} 3000 \gauss for $\alpha B_R$, $\alpha B_\phi$, $\alpha B_\theta$ following \autoref{fig:fig1_qualitative}.
    Data: 2016 January 26, 23:12 TAI.}
    \label{fig:realizations}
\end{figure*}

We begin with qualitative comparison of these realizations to determine where magnetic structures remain consistent and where stochastic variations occur.
\autoref{fig:realizations} shows three independent realizations drawn from the same UV/EUV observations. The realizations recover almost identical large scale magnetic structures and polarity across \abr, \abp, and \abt, indicating that the dominant magnetic structures are well constrained. Meanwhile, small differences appear in localized weak-field regions, reflecting the stochastic nature of diffusion sampling process. This behavior suggests that \modelname does not produce arbitrary samples, but instead generates multiple plausible magnetograms that remain consistent in well-constrained regions while allowing variation when the signal is weak or uncertain.

\subsubsection{Quantitative Comparison between Realizations}

Given this constrained variation across realizations, we next use a best-of-$k$ analysis to test whether drawing more samples increases the chance of obtaining a realization that more closely matches the \hinode ground truth. \autoref{tab:bestofn} reports this evaluation using the same metrics and evaluation method as \autoref{tab:summary}. For each sample in the test set, we draw $k$ independent \modelname realizations and evaluate the realization with smallest error relative to the \hinode ground truth.

This best-of-$k$ evaluation is not intended to represent normal inference, where the ground truth is unavailable. Rather, it is a controlled test of the learned predictive distribution: if \modelname places probability mass near the ground truth magnetogram, then drawing more samples should increase the chance of obtaining a realization closer to the \hinode ground truth. This trend is confirmed in \autoref{tab:bestofn}: increasing $k$ leads to consistent improvements in pixel-wise accuracy across all field components.
In contrast, the distributional and structural metrics remain largely unchanged. These results indicate that additional sampling can increase the possibility of obtaining a realization that has better per-pixel correspondence with the \hinode observation while keeping the overall pixel-value distribution, sharpness, and spectral characteristics. Together with \autoref{fig:realizations}, these results show that stochastic variations among realizations permit generalization of multiple slightly different yet distributionally and structurally constrained magnetograms.

\begin{deluxetable*}{lc@{\,}c@{\,}c@{\,}c@{\,}c@{\,}c@{\,}cc@{\,}c@{\,}c@{\,}c@{\,}c@{\,}c@{\,}cc@{\,}c@{\,}c@{\,}c@{\,}c@{\,}c@{\,}cc@{\,}c@{\,}c@{\,}c@{\,}c@{\,}c@{\,}c}
\tablewidth{0pt}
\tabletypesize{\scriptsize}
\tabcolsep=5pt
\tablecaption{\textbf{Best-of-$N$ selection improves \modelname's per-pixel accuracy with diminishing returns, while a single realization already attains its full structural fidelity.}
We evaluate \modelname on the test set when multiple attempts are permitted: for each input, \modelname samples $N$ candidate realizations, and we evaluate the candidate with the smallest error against the \hinode ground truth. Each cell lists best-of-$N$ for $N = 1, 4, 16, 32$, with the best value in \textbf{bold}.
Metrics follow \autoref{tab:summary}: for each field component, MAE (lower is better) and \%$<300$ (higher is better) are computed over strong-field pixels with $|\alpha \mathbf{B}| > 1000$~\gauss, while $W_1$, the $\nabla$-ratio, $\nabla$-MAE, and RALSD measure how well predictions reproduce the pixel-value distribution, sharpness, and power spectrum of the ground truth over all valid pixels.
\label{tab:bestofn}}
\tablehead{\multicolumn{1}{l}{Metric} & \multicolumn{7}{c}{\abr} & \multicolumn{7}{c}{\abp} & \multicolumn{7}{c}{\abt} & \multicolumn{7}{c}{$|\alpha\BB|$}}
\startdata
    \multicolumn{1}{l}{MAE [\gauss] $\downarrow$} & 575.5 & $|$ & 552.1 & $|$ & 538.4 & $|$ & \textbf{533.1} & 304.3 & $|$ & 297.5 & $|$ & 293.3 & $|$ & \textbf{291.2} & 294.6 & $|$ & 286.3 & $|$ & 282.0 & $|$ & \textbf{280.0} & 372.8 & $|$ & 371.7 & $|$ & 370.9 & $|$ & \textbf{370.7} \\
    \multicolumn{1}{l}{\%$<$300 $\uparrow$} & 49.1 & $|$ & 49.9 & $|$ & 50.4 & $|$ & \textbf{50.6} & 69.8 & $|$ & 70.5 & $|$ & 71.0 & $|$ & \textbf{71.2} & 71.3 & $|$ & 72.1 & $|$ & 72.5 & $|$ & \textbf{72.7} & 57.0 & $|$ & 57.2 & $|$ & 57.3 & $|$ & \textbf{57.4} \\
    \multicolumn{1}{l}{$W_1$ [\gauss] $\downarrow$} & \textbf{14.9} & $|$ & 15.0 & $|$ & 15.0 & $|$ & 15.1 & 10.5 & $|$ & 10.5 & $|$ & 10.5 & $|$ & 10.5 & 10.9 & $|$ & 10.9 & $|$ & 10.9 & $|$ & 10.9 & \textbf{24.7} & $|$ & 24.8 & $|$ & 24.8 & $|$ & 24.9 \\
    \multicolumn{1}{l}{$\nabla$-ratio $\rightarrow\!1$} & 0.73 & $|$ & 0.73 & $|$ & 0.73 & $|$ & 0.73 & \textbf{0.62} & $|$ & 0.61 & $|$ & 0.61 & $|$ & 0.61 & 0.60 & $|$ & 0.60 & $|$ & 0.60 & $|$ & 0.60 & 0.71 & $|$ & 0.71 & $|$ & 0.71 & $|$ & 0.71 \\
    \multicolumn{1}{l}{$\nabla$-MAE [\gauss/px] $\downarrow$} & 38.05 & $|$ & 38.01 & $|$ & 38.00 & $|$ & \textbf{37.99} & 20.93 & $|$ & 20.92 & $|$ & 20.91 & $|$ & \textbf{20.91} & 20.96 & $|$ & 20.95 & $|$ & 20.94 & $|$ & \textbf{20.94} & 34.72 & $|$ & 34.68 & $|$ & 34.67 & $|$ & \textbf{34.66} \\
    \multicolumn{1}{l}{RALSD [dB] $\downarrow$} & 4.25 & $|$ & \textbf{4.24} & $|$ & 4.24 & $|$ & 4.24 & 6.00 & $|$ & 6.00 & $|$ & 6.00 & $|$ & 6.00 & \textbf{5.96} & $|$ & 5.97 & $|$ & 5.97 & $|$ & 5.97 & \textbf{4.04} & $|$ & 4.04 & $|$ & 4.05 & $|$ & 4.04 \\
\enddata
\end{deluxetable*}
\onecolumngrid

\subsection{Additional Full-Disk Quantitative Results}

\begin{figure*}[t]
    \centering
    \includegraphics[width=1\linewidth]{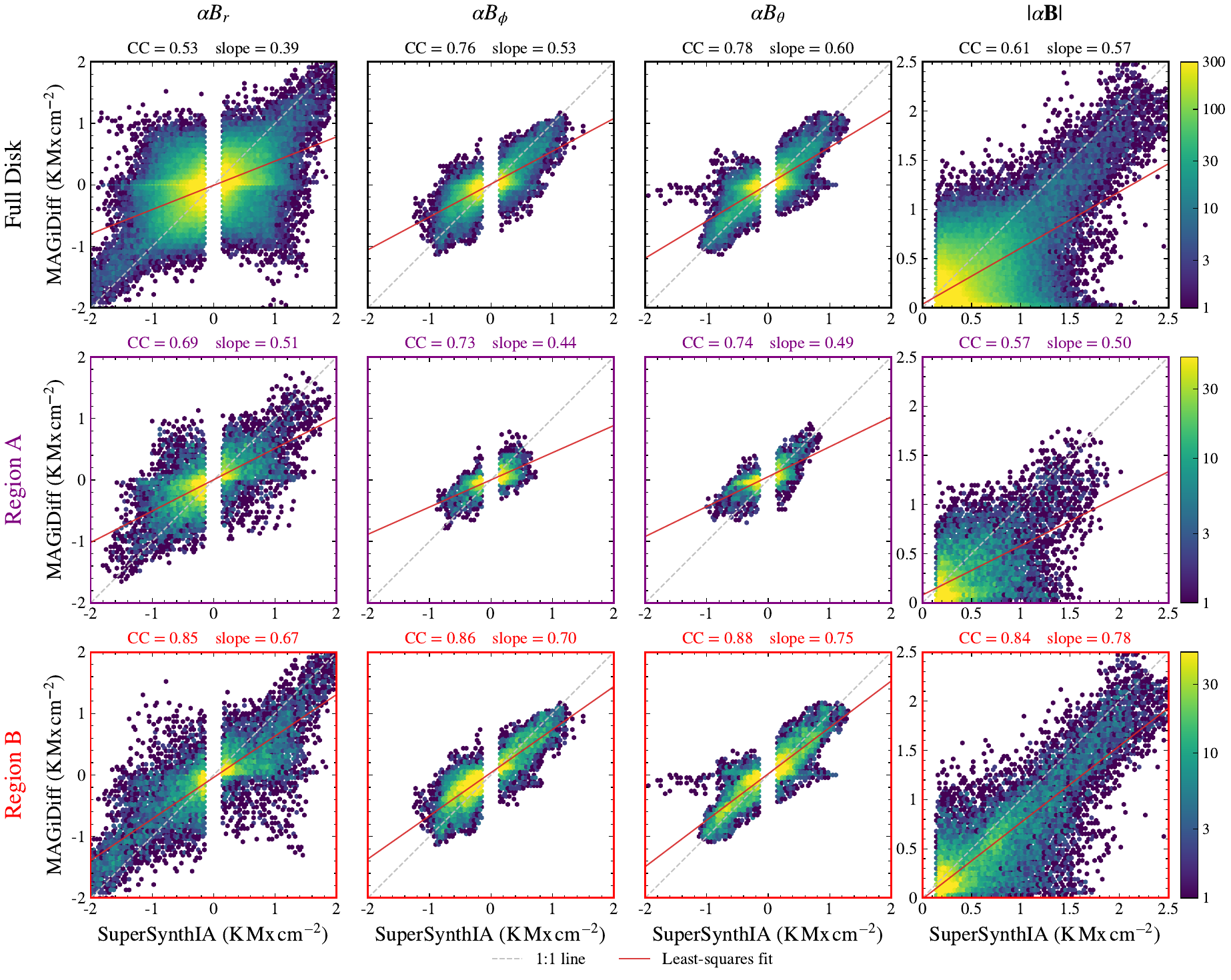}
     \caption{\textbf{Hexbin density plots of \modelname predictions against SuperSynthIA vector magnetograms for the full-disk sample shown in \autoref{fig:full_disk}. } 
     From left to right, each column corresponds to \abr, \abp, \abt, and $|\alpha \mathbf{B}|$. From top to bottom, each row shows the hexbin plots for the full-disk, region A, and region B, respectively. 
     Each panel shows the hexbin density together with a gray dashed line marking the $1{:}1$ relation and a red line marking the least-squares fit on all pixels. We also report the Pearson correlation coefficient (CC) and the slope of the fit line on top of each panel. 
     Most of the density concentrates near the $1{:}1$ line, indicating strong agreement between \modelname prediction and the reference SuperSynthIA magnetogram, with the strongest agreement observed in the most active region B. Pixels with ground truth absolute value below $150$~\gauss are omitted from all panels to make trends more apparent.
     } 
    \label{fig:full_disk_hexbin}
\end{figure*}

To quantitatively evaluate \modelname on the full-disk sample shown in \autoref{fig:full_disk}, \autoref{fig:full_disk_hexbin} presents hexbin plots comparing \modelname predictions with SuperSynthIA vector magnetograms for the full disk and for region A and B separately.

\subsection{Cross-Instrument Fine-Tuning Pipeline}
\label{sec:appx_cross_instr_fine_tuning}

To adapt \modelname to EUV instruments other than \aia, we keep the learning problem fixed and handle instrumental differences during preprocessing. This preprocessing aims to bring each instrument's EUV filtergrams into closer agreement with \aia filtergrams before fine-tuning. For each instrument, we first reproject all filtergrams onto a common grid and resample them to reduce pixel-resolution mismatch with \aia. We then calibrate the intensities to approximately the \aia range and set unavailable \aia channels to zero.
The fine-tuned model therefore receives the same conditioning input: \aia-like EUV filtergrams, an on-disk mask, a heliographic latitude map, and a solar cycle indicator. 

Since \hinode does not co-observe \stereo or \goes densely enough for paired training, we use SuperSynthIA predictions as the target vector magnetograms. For each EUV observation, we take the nearest SuperSynthIA \abr, \abp, \abt maps and reproject them onto the corresponding EUV filtergram grid. Because SuperSynthIA follows \hmi \abt sign convention, a flip on SuperSynthIA \abt sign is necessary after reprojection to match the \hinode convention used by the pretrained \modelname.

Fine-tuning proceeds in two stages. We first fine-tune the VAE on $256 \times 256$ pixel SuperSynthIA vector magnetogram crops so that its latent space is adapted to SuperSynthIA predictions rather than \hinode magnetograms. This corresponds to approximately $153\arcsec \times 153 \arcsec$. We then load the pretrained denoising U-Net, replace the original VAE with the SuperSynthIA fine-tuned VAE, reset the optimizer, and fine-tune on instrument-specific pairs of EUV filtergrams and target magnetograms using $512 \times 512$ pixel crops, corresponding to approximately $306\arcsec \times 306\arcsec$, while keeping the VAE frozen. Otherwise, the denoising training setup is unchanged.

During fine-tuning, we also use a sign-flip augmentation analogous to the one used in the cross solar cycle generalization test in Section~\ref{sec:cross_sc}. With probability $0.5$, the target vector magnetogram is multiplied by $-1$ and the solar cycle indicator is changed from the current cycle parity to the opposite parity, while the EUV filtergrams and latitude map remain unchanged. This augmentation is included because instrument-specific fine-tuning datasets are relatively small and may not span multiple solar cycles. The remaining preprocessing is instrument specific.

\subsubsection{Adaption to \stereo}
\label{sec:appx_migration_to_stereo}

Intuitively, directly applying \modelname trained on \aia data to \stereo filtergrams will not work due to instrumental differences in field of view, optical resolution, calibration, and degradation. We therefore make several adjustments before fine-tuning. We use data from 2022 to 2024 to construct the fine-tuning dataset because during this period \stereoA is near Earth and has substantial field-of-view overlap with \hmi. This overlap is essential because we use SuperSynthIA predictions as the target vector magnetograms, and \hmi stokes observations are the input to SuperSynthIA. 

For the fine-tuning dataset, we use $171$\AA, $195$\AA, and $304$\AA~ channels because they are the \stereo passbands that overlap with \aia passbands. We sample candidate times at a $4$-hour cadence and, because these channels are observed asynchronously on \stereo,  retain only timestamps for which all three channels are available within $5$ minutes. Raw Level-$0.5$ FITS files are calibrated to Level-$1$ with SECCHI\_PREP pipeline in SolarSoft package \citep{solarsoft_1998}, which applies bias subtraction, flat-field correction, and are converted to physical units \DNS. We then reproject the $195$\AA~ and $304$\AA~ maps onto the $171$\AA~ coordinate frame and bilinearly upsample the resulting maps by a factor of $2.67\times$ to match \aia's angular pixel scale. We then reproject the SuperSynthIA vector magnetograms onto this upsampled grid. After filtering for valid paired full-disk EUV filtergrams and SuperSynthIA vector magnetograms, the \stereo fine-tuning dataset contains $4949$ pairs. We split these data temporally: $2022$ January through $2023$ June for training ($2938$ pairs), $2023$ July through December for validation ($523$ pairs), and $2024$ April through December for testing ($1488$ pairs).

\begin{deluxetable*}{lc@{\,}c@{\,}c@{\,}c@{\,}c@{\,}c@{\,}cc@{\,}c@{\,}c@{\,}c@{\,}c@{\,}c@{\,}cc@{\,}c@{\,}c@{\,}c@{\,}c@{\,}c@{\,}cc@{\,}c@{\,}c@{\,}c@{\,}c@{\,}c@{\,}c}
\tablewidth{0pt}
\tabletypesize{\scriptsize}
\tabcolsep=5pt
\tablecaption{\textbf{On the more challenging \stereo cutouts, the fine-tuned \modelname still produces reasonable estimates, and best-of-$k$ sampling improves performance.} Evaluation setup follows \autoref{tab:bestofn}.
\label{tab:appx_ft_stereo} }
\tablehead{\multicolumn{1}{l}{Metric} & \multicolumn{7}{c}{\abr} & \multicolumn{7}{c}{\abp} & \multicolumn{7}{c}{\abt} & \multicolumn{7}{c}{$|\alpha\BB|$}}
\startdata
    \multicolumn{1}{l}{MAE [\gauss] $\downarrow$} & 779.0 & $|$ & 726.9 & $|$ & 692.1 & $|$ & \textbf{677.6} & 355.6 & $|$ & 335.3 & $|$ & 320.4 & $|$ & \textbf{315.6} & 354.7 & $|$ & 333.0 & $|$ & 317.1 & $|$ & \textbf{310.6} & 786.3 & $|$ & 739.9 & $|$ & 702.7 & $|$ & \textbf{687.4} \\
    \multicolumn{1}{l}{\%$<$300 $\uparrow$} & 19.9 & $|$ & 22.3 & $|$ & 24.2 & $|$ & \textbf{25.1} & 51.6 & $|$ & 54.1 & $|$ & 56.3 & $|$ & \textbf{56.9} & 52.0 & $|$ & 54.7 & $|$ & 56.9 & $|$ & \textbf{57.9} & 19.5 & $|$ & 22.0 & $|$ & 24.0 & $|$ & \textbf{25.0} \\
    \multicolumn{1}{l}{$W_1$ [\gauss] $\downarrow$} & 9.2 & $|$ & 9.3 & $|$ & 9.2 & $|$ & \textbf{9.2} & 4.6 & $|$ & 4.7 & $|$ & 4.6 & $|$ & \textbf{4.5} & 4.5 & $|$ & 4.5 & $|$ & 4.5 & $|$ & \textbf{4.4} & 12.7 & $|$ & 12.8 & $|$ & 12.7 & $|$ & \textbf{12.6} \\
    \multicolumn{1}{l}{$\nabla$-ratio $\rightarrow\!1$} & 0.75 & $|$ & 0.75 & $|$ & 0.75 & $|$ & 0.75 & 0.70 & $|$ & 0.70 & $|$ & 0.70 & $|$ & 0.70 & \textbf{0.70} & $|$ & 0.69 & $|$ & 0.69 & $|$ & 0.69 & \textbf{0.76} & $|$ & 0.75 & $|$ & 0.75 & $|$ & 0.75 \\
    \multicolumn{1}{l}{$\nabla$-MAE [\gauss/px] $\downarrow$} & 18.40 & $|$ & 18.28 & $|$ & 18.20 & $|$ & \textbf{18.16} & 5.27 & $|$ & 5.21 & $|$ & 5.17 & $|$ & \textbf{5.15} & 5.51 & $|$ & 5.45 & $|$ & 5.40 & $|$ & \textbf{5.39} & 18.03 & $|$ & 17.91 & $|$ & 17.83 & $|$ & \textbf{17.79} \\
    \multicolumn{1}{l}{RALSD [dB] $\downarrow$} & 3.93 & $|$ & \textbf{3.93} & $|$ & 3.94 & $|$ & 3.96 & 4.73 & $|$ & 4.71 & $|$ & 4.70 & $|$ & \textbf{4.70} & 4.16 & $|$ & 4.14 & $|$ & \textbf{4.14} & $|$ & 4.15 & 3.85 & $|$ & \textbf{3.85} & $|$ & 3.86 & $|$ & 3.88 \\
\enddata
\end{deluxetable*}
\onecolumngrid

The main \stereo-specific challenge is photometric mismatch. The paired \stereo--\aia intensity distributions are not well matched by a single affine correction, especially in the bright active-region and low-intensity tails. This mismatch is further complicated by long-term \stereo degradation: the SECCHI\_PREP pipeline calibrates the raw data to Level-1, but it does not fully remove long-term changes in instrument sensitivity, so the \stereo intensity distribution can vary substantially over the mission. We therefore use a time-varying quantile mapping in $\log(1+x)$ space. The mapping is estimated from about $10$ calibration pairs per available month from $2011$ through $2025$. Each pair consists of \stereo and \aia observations taken close in time. We exclude September 2014 through November 2015 because \stereoA was in solar conjunction and data are unavailable. This leaves $1606$ calibration pairs. For each pair, the on-disk pixels are sorted by intensity separately for \stereo and \aia. The $0$th, $1$st, \ldots, $100$th percentiles of these sorted intensity values are then recorded for each \stereo-to-aia channel pair (\stereo $171$\AA $\rightarrow$ \aia $171$\AA, \stereo $195$\AA $\rightarrow$ \aia $193$\AA, and \stereo $304$\AA $\rightarrow$ \aia $304$\AA). These percentiles form the raw calibration anchors for the lookup table by matching intensity values at the same percentile rank in the two instruments. For example, if the median on-disk \stereo $195$\AA~ intensity in a calibration pair is $x$ and the median on-disk \aia $193$\AA ~ intensity in the corresponding \aia filtergram is $y$, then the $50$th-percentile anchor maps $x \mapsto y$. The same procedure is applied to all percentiles, so the mapping matches the full intensity distribution. 

The calibration anchors are grouped into semiannual time bins. Within each bin and each channel pair, the matched percentile values are median-aggregated to form a stable lookup curve from \stereo intensity to \aia intensity. The curves are smoothed with a three-bin rolling median, and monotonicity is enforced. This produces a lookup curve with $101$ breakpoints. 

To map a \stereo intensity value, we first apply $\log(1+x)$ and then linearly interpolate between the neighboring breakpoints, using linear extrapolation for values outside the stored breakpoint range. This places the \stereo values on the same $\log(1+x)$ intensity scale as the \aia inputs used to pretrain \modelname. After this transform, training proceeds with standard \aia normalization.

This correction reduces the dominant distribution shift associated \stereo degradation, but it does not make the model equally reliable at all observation times. Because the fine-tuning dataset is restricted to observation from $2022$ through $2024$, when \stereoA has substantial overlap with \hmi, we expect the fine-tuned \modelname model to be most reliable near this time range. Applying the model to much earlier \stereo observations may therefore require additional adjustment, either by extending the fine-tuning dataset to earlier observations or by improving the calibration for long-term \stereo intensity changes.

We quantify the fine-tuned \modelname performance on the held out test set in \autoref{tab:appx_ft_stereo}, using SuperSynthIA as the reference.

\subsubsection{Adaption to \goes}
\label{sec:appx_migration_to_goes}

\begin{deluxetable*}{lc@{\,}c@{\,}c@{\,}c@{\,}c@{\,}c@{\,}cc@{\,}c@{\,}c@{\,}c@{\,}c@{\,}c@{\,}cc@{\,}c@{\,}c@{\,}c@{\,}c@{\,}c@{\,}cc@{\,}c@{\,}c@{\,}c@{\,}c@{\,}c@{\,}c}
\tablewidth{0pt}
\tabletypesize{\scriptsize}
\tabcolsep=5pt
\tablecaption{\textbf{On strong-field \goes cutouts, the fine-tuned \modelname produces reasonable estimates, with best-of-$k$ sampling improves performance.} Evaluation setup follow \autoref{tab:bestofn}. 
\label{tab:appx_ft_goes} }
\tablehead{\multicolumn{1}{l}{Metric} & \multicolumn{7}{c}{\abr} & \multicolumn{7}{c}{\abp} & \multicolumn{7}{c}{\abt} & \multicolumn{7}{c}{$|\alpha\BB|$}}
\startdata
    \multicolumn{1}{l}{MAE [\gauss] $\downarrow$} & 805.0 & $|$ & 747.6 & $|$ & 709.4 & $|$ & \textbf{695.9} & 348.4 & $|$ & 326.0 & $|$ & 311.3 & $|$ & \textbf{306.2} & 354.3 & $|$ & 331.4 & $|$ & 315.8 & $|$ & \textbf{309.9} & 791.7 & $|$ & 750.9 & $|$ & 717.3 & $|$ & \textbf{704.2} \\
    \multicolumn{1}{l}{\%$<$300 $\uparrow$} & 19.2 & $|$ & 21.5 & $|$ & 23.4 & $|$ & \textbf{24.1} & 53.0 & $|$ & 55.7 & $|$ & 57.7 & $|$ & \textbf{58.5} & 52.3 & $|$ & 54.9 & $|$ & 57.1 & $|$ & \textbf{57.9} & 19.3 & $|$ & 21.4 & $|$ & 23.3 & $|$ & \textbf{24.1} \\
    \multicolumn{1}{l}{$W_1$ [\gauss] $\downarrow$} & 8.1 & $|$ & 8.1 & $|$ & 8.1 & $|$ & \textbf{8.0} & 4.2 & $|$ & 4.3 & $|$ & 4.2 & $|$ & \textbf{4.2} & 4.1 & $|$ & 4.1 & $|$ & 4.1 & $|$ & \textbf{4.0} & 11.4 & $|$ & 11.5 & $|$ & 11.4 & $|$ & \textbf{11.3} \\
    \multicolumn{1}{l}{$\nabla$-ratio $\rightarrow\!1$} & 0.78 & $|$ & 0.78 & $|$ & 0.78 & $|$ & 0.78 & \textbf{0.72} & $|$ & 0.71 & $|$ & 0.71 & $|$ & 0.71 & \textbf{0.72} & $|$ & 0.72 & $|$ & 0.71 & $|$ & 0.72 & \textbf{0.79} & $|$ & 0.78 & $|$ & 0.78 & $|$ & 0.78 \\
    \multicolumn{1}{l}{$\nabla$-MAE [\gauss/px] $\downarrow$} & 19.12 & $|$ & 18.99 & $|$ & 18.90 & $|$ & \textbf{18.86} & 5.37 & $|$ & 5.32 & $|$ & 5.27 & $|$ & \textbf{5.26} & 5.62 & $|$ & 5.56 & $|$ & 5.51 & $|$ & \textbf{5.50} & 18.73 & $|$ & 18.61 & $|$ & 18.51 & $|$ & \textbf{18.48} \\
    \multicolumn{1}{l}{RALSD [dB] $\downarrow$} & \textbf{2.72} & $|$ & 2.73 & $|$ & 2.72 & $|$ & 2.72 & \textbf{3.70} & $|$ & 3.73 & $|$ & 3.72 & $|$ & 3.72 & 3.65 & $|$ & 3.65 & $|$ & 3.65 & $|$ & \textbf{3.64} & \textbf{2.69} & $|$ & 2.70 & $|$ & 2.70 & $|$ & 2.69 \\
\enddata
\end{deluxetable*}
\onecolumngrid

As a second cross-instrument adaption test, we fine-tune \modelname on the Solar Ultraviolet Imager (SUVI) onboard \goessixteen and \goeseighteen.

For the fine-tuning dataset, we use $94$\AA, $131$\AA, $171$\AA, $195$\AA, and $304$\AA~ channels since they are the passbands that overlap with \aia passbands. We sample the Level-2 FITS products at a $4$-hour cadence and require all selected channels to be observed within $10$ minutes from each other. For each timestamp, the five \goes filtergrams are coaligned to the $171$\AA~ coordinate frame and then bilinearly upsampled by $4.17\times$ to match \aia's angular pixel scale. We then reproject the SuperSynthIA vector magnetograms onto the upsampled \goes grid. 

The fine-tuning dataset spans $2022$ through $2024$ and contains $7475$ data pairs across \goessixteen and \goeseighteen. After filtering, the train/validation/test datasets contain $3160$/$1231$/$2208$ samples. As in the \stereo fine-tuning experiment, we split the data temporally: $2022$ January through $2023$ June for training, $2023$ July through December for validation, and $2024$ April through December for testing.

We use the calibrated Level-2 products directly and normalize the inputs in $\log(1+x)$ space. Since the two satellites are not fully intercalibrated, we compute separate per-channel means and standard deviations for \goessixteen and \goeseighteen. Fine-tuning is then performed on combined data pairs from both satellites, with each sample normalized using the parameters for its source satellite.

The fine-tuned \modelname performance on the held out test set is evaluated in \autoref{tab:appx_ft_goes}, using SuperSynthIA as the reference.

\subsubsection{Additional Quantitative Results}

\begin{figure*}[t]
    \centering
    \includegraphics[width=1\linewidth]{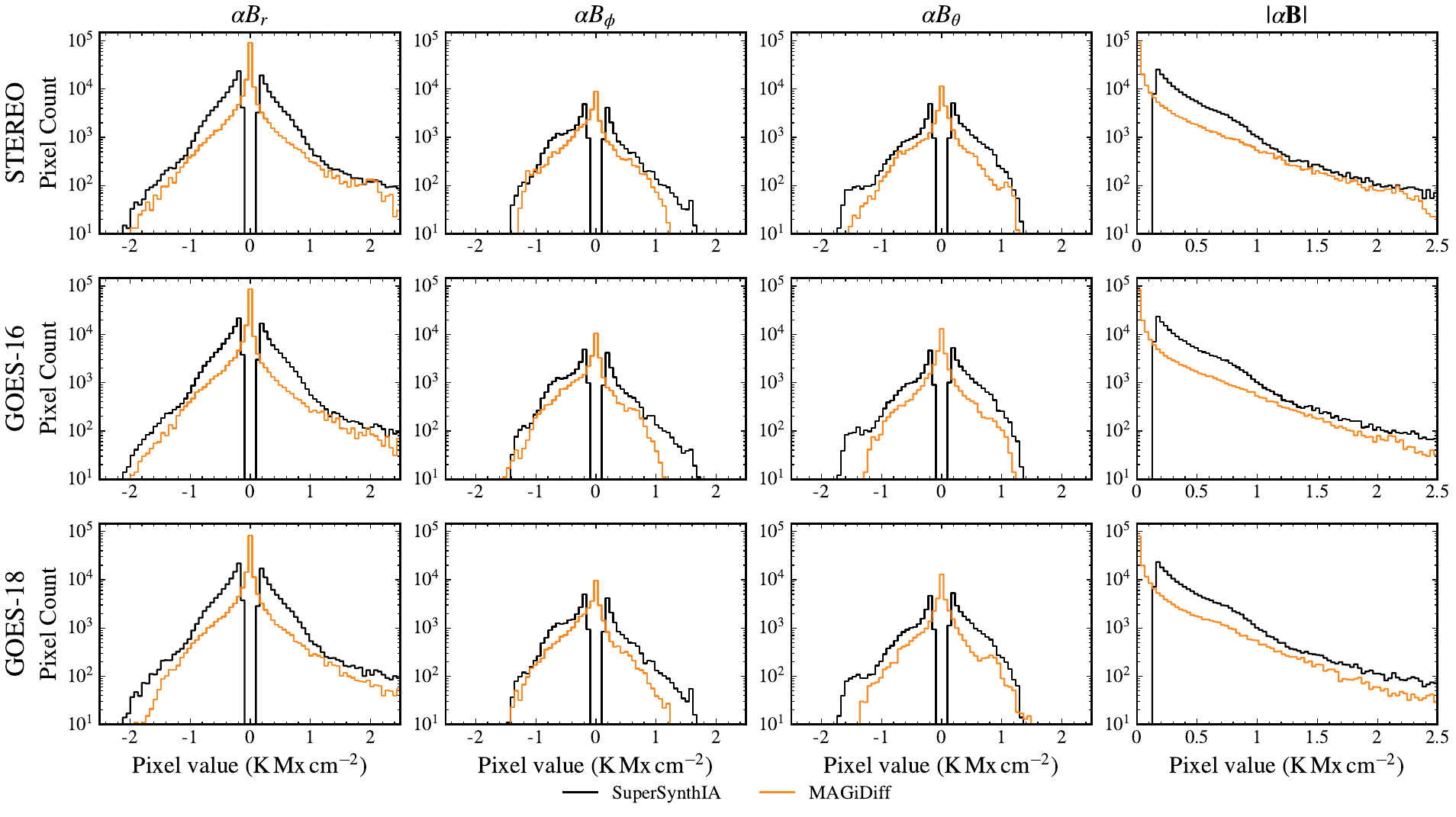}
     \caption{\textbf{Pixel-value histograms comparing \modelname predictions with the corresponding SuperSynthIA vector magnetograms for the full-disk observation shown in \autoref{fig:other_Instr_full_disk}. } 
     From left to right, each column correspond to \abr, \abp, \abt, and $|\alpha \mathbf{B}|$. From top to bottom, the row correspond to full-disk \stereo, \goessixteen, and \goeseighteen predictions, respectively. } 
    \label{fig:other_Instr_full_disk_hist}
\end{figure*}

\autoref{fig:other_Instr_full_disk_hist} provides an additional quantitative analysis of the full-disk predictions shown in \autoref{fig:other_Instr_full_disk}, comparing the \modelname prediction distributions with those of the corresponding reference SuperSynthIA magnetograms.

\end{document}